\documentclass{elsart}
\usepackage{latexsym}
\usepackage{amsmath}
\usepackage[dvips]{graphicx}
\newcommand{\beq}{\begin{equation}}
\newcommand{\eeq}{\end{equation}}
\newcommand{\beqa}{\begin{eqnarray}}
\newcommand{\eeqa}{\end{eqnarray}}
\newcommand{\bseq}{\begin{subequations}}
\newcommand{\eseq}{\end{subequations}}
\newcommand{\bold}{\boldsymbol}
\newcommand{\wtilde}{\widetilde}

\newcommand{\trm}{\textrm}

\begin{document}

\begin{frontmatter}

\title{Finite momentum meson correlation functions in a QCD plasma}
\author{W.M. Alberico$^1$, A. Beraudo$^1$, P. Czerski$^2$ and A. Molinari$^1$}, 
\address{$^1$Dipartimento di Fisica Teorica dell'Universit\`a di Torino and \\ 
  Istituto Nazionale di Fisica Nucleare, Sezione di Torino, \\ 
  via P.Giuria 1, I-10125 Torino, Italy\\
  \ \\
  $^2$ Institute of Nuclear Physics Polish Academy of Science,
  Krak\'ow,\\
  ul. Radzikowskiego 152, Poland}

%\date{\today}

\begin{abstract}
The finite momentum meson spectral function (MSF) in the pseudoscalar channel is evaluated, adopting for the 
fermionic propagators HTL expressions. The different contributions to the meson spectral functions are 
clearly displayed. 
Our analysis may be of relevance for lattice studies of MSF based so far on the Maximum Entropy Method.
As a further step the correlation function along the (imaginary-) temporal direction is evaluated. 
\end{abstract}
\end{frontmatter}
\section{Introduction}
In a previous paper \cite{berry} we studied the zero momentum Meson Spectral Functions (MSFs) and the correlations along the (imaginary) temporal direction in the QGP phase.\\
Our starting point was the calculation done in Ref.~\cite{mus}, where the above quantities were evaluated in the Hard Thermal Loop (HTL) approximation.\\   
The main result of Ref.~\cite{mus} was the unraveling of the behavior displayed by the MSF at small values of the energy $\omega$. Here indeed, owing to the presence of the plasmino mode in the HTL quark propagator, divergences appear in the density of states giving rise to peaks in the MSF, referred to as \emph{Van Hove singularities}.\\
The possible experimental relevance of such singularities was analyzed in Ref.~\cite{bra}, where the back-to-back dilepton production rate was evaluated in terms of the zero momentum MSF in the vector channel.\\
Actually in Ref.~\cite{mus} HTL resummed propagators, in principle valid only for the soft modes, were adopted over the whole range of momenta.\\
Hence in our previous paper \cite{berry} we compared these results with those obtained with a more realistic  treatment of the hard fermionic modes. We found that the qualitative behavior of the MSF for small values of $\omega$, arising from the presence of two branches (normal quark mode and plasmino) plus a continuum part in the HTL quark spectral function, was mildly affected.\\

In the present paper we extend the analysis to the case of finite spatial momentum. Since the problem turns out to be numerically more involved and since, as above stated, the qualitative structure of the MSF is substantially unchanged when corrections to the HTL scheme are accounted for, here we use HTL resummed fermionic propagators over the whole range of integration in momentum space.\\

The analytical study of the spatial mesonic correlations was first addressed, in the non interacting case, in Ref.~\cite{flor}.\\
The study of these correlations is of particular relevance in finite temperature lattice QCD. The large number of lattice sites available along the spatial directions allows one to study the large distance behavior of the mesonic correlators. This displays an exponential decay governed by a \emph{screening mass}. Results obtained on the lattice can be found, for example, in Refs.~\cite{taro,taro2,petr,wis}. Concerning analytical approaches, effects of the interaction on the meson screening masses were analyzed in Refs.~\cite{za,lai} within a dimensional reduction framework.\\
The knowledge of the screening masses leads to the understanding of the nature of the excitations characterizing the deconfined phase of QCD at the different temperatures: weakly interacting quarks and gluons or strongly correlated states.\\
In this paper we confine ourselves to the study of finite momentum mesonic correlators, leaving the spatial correlations (along the z-axis) for future work.\\  
 
Indeed the full information on the momentum and temperature behavior of the mesonic excitations in the QGP phase is encoded in their spectral function. Unfortunately the latter is not measured directly on the lattice, but has to be reconstructed from the temporal correlator, which is known for a rather small set of points. This is the reason why most works are confined to the evaluation of the screening masses, even if the information which can be extracted from the latter is much poorer with respect to the one stemming from the MSF. Nevertheless some results for the finite momentum MSF (for light mesons) start to be available \cite{aar,aar2}.\\
Presently the MSFs are extracted from the temporal correlators through a technique called Maximum Entropy Method (MEM). For a review of this approach we refer the reader to \cite{mem}.\\
We notice that, beside the study of meson properties at finite temperature, the problem of extracting spectral densities from lattice euclidean correlators is encountered in the evaluation of other quantities of interest for QGP phenomenology like the thermal dilepton production rate \cite{dilep}, the soft photon emission and the electrical conductivity \cite{gupta}, the hydrodynamical transport coefficients (shear and bulk viscosity and heat conductivity) \cite{naka,aar3,wyld}. The link between euclidean correlators, spectral densities and transport coefficients is provided, in the framework of linear response theory, by the Kubo relations \cite{kubo1,kubo2,kubo3}.\\  
Lattice results for finite momentum MSF are presently getting available also in the heavy meson sector. In particular the finite momentum charmonium spectral function (describing a $c\bar{c}$ meson moving in the heat bath frame) is found to displays a peaked structure in the pseudoscalar ($\eta_c$) and vector ($J/\Psi$) channels even at temperatures above $T_c$ \cite{dat}.\\
Since a $J/\Psi$ moving in the heat bath frame sees more energetic gluons than at rest, the occurrence of a well defined peak in its spectral function seems to point to a robust binding of such a state since the collisions with ``blue-shifted'' gluons do not suffice to melt it.\\
We hope that our results can provide a useful starting point or testing ground for MEM studies of MSF.

Our paper is organized as follows.\\
We shortly review in Sec. \ref{sec:mesons} the basic definitions of the mesonic operators, correlators and spectral functions addressed in this work.\\
In Sec. \ref{sec:quark} we briefly discuss the results of the HTL approximation for the quark propagator.\\
Sec. \ref{sec:pseudosc} represents the central part of our paper. We focus our attention on the pseudoscalar channel. In this Section we present our analytical and numerical results both for the MSF (Secs. \ref{sec:spec} and \ref{spec_num}) and for the temporal correlator (Sec. \ref{sec:tempor}).\\
Finally in Sec. \ref{sec:concl} we summarize and discuss our findings.
 
\section{Finite temperature meson spectral functions}\label{sec:mesons}
Following the notation adopted in Ref. \cite{mus} we consider the following current operator,
 carrying the quantum numbers of a meson:
\beq
J_M(-i\tau,\bold{x})=\bar{q}(-i\tau,\bold{x})\Gamma_M q(-i\tau,\bold{x})\;,
\eeq
where $\Gamma_M=1,\gamma^5,\gamma^{\mu},\gamma^{\mu}\gamma^5$ for the scalar, pseudoscalar, vector
and pseudo vector channels, respectively. We next define the fluctuation operator $\wtilde{J}_M$ as
\beq
\wtilde{J}_M(-i\tau,\bold{x})=J_M(-i\tau,\bold{x})-\langle J_M(-i\tau,\bold{x})\rangle\;,
\eeq
the average being taken on the grand canonical ensemble.\\
The chief quantity we address is the \textit{thermal  meson 2 point correlation function}:
\beqa
\chi_M(-i\tau,\bold{x}) & = & \langle \wtilde{J}_M(-i\tau,\bold{x})\wtilde{J}_M^{\dagger}(0,\bold{0})
\rangle \nonumber\\
{} & = & \langle J_M(-i\tau,\bold{x})J_M^{\dagger}(0,\bold{0})\rangle-\langle J_M(-i\tau,\bold{x})
\rangle\langle J_M^{\dagger}(0,\bold{0})\rangle\nonumber\\
{} & = & \frac{1}{\beta}\sum_{n=-\infty}^{+\infty}\int\frac{d^3p}{(2\pi)^3}e^{-i\omega_n\tau}
e^{i\bold{p}\cdot\bold{x}}\chi_M(i\omega_n,\bold{p})\;,\label{eq:mescorr}
\eeqa
with $\tau\in[0,\beta=1/T]$ and $\omega_n=2n\pi T$ ($n=0,\pm1,\pm2\dots$).\\
It is convenient to use the following spectral representation for the meson propagator in momentum space:
\beq
\chi_{M}(i\omega_n,\bold{p})=-\int\limits_{-\infty}^{+\infty}d\omega\frac{\sigma_{M}(\omega,\bold{p})}
{i\omega_n-\omega} \quad \Rightarrow \quad \sigma_{M}(\omega,\bold{p})=
\frac{1}{\pi}\trm{Im}\,\chi_{M}(\omega+i\eta,\bold{p}).\label{eq:specrit}
\eeq
Hence it is possible to express the thermal meson propagator in a mixed representation through the
 \textit{spectral function} $\sigma_{M}$. Indeed starting from the definition:
\beqa
G_{M}(-i\tau,\bold{p})& = & \frac{1}{\beta}\sum_{n=-\infty}^{+\infty}e^{-i\omega_n\tau}
\chi_M(i\omega_n,\bold{p}) \nonumber\\
{} & = & -\frac{1}{\beta}\sum_{n=-\infty}^{+\infty}e^{-i\omega_n\tau}\int\limits_{-\infty}^{+\infty}
d\omega\frac{\sigma_{M}(\omega,\bold{p})}{i\omega_n-\omega}\;,\nonumber
\eeqa
and performing the sum over the Matsubara frequencies with a standard contour integration in
the complex $\omega$ plane \cite{lb,bi}, one obtains \cite{mus}:
\beq
G_{M}(-i\tau,\bold{p})=\int\limits_0^{+\infty}d\omega~\sigma_{M}(\omega,\bold{p})
\frac{\cosh[\omega(\tau-\beta/2)]}{\sinh(\omega\beta/2)}\equiv \int\limits_0^{+\infty}d\omega~\sigma_{M}
(\omega,\bold{p})K(\omega,\tau).
\label{eq:gtau}
\eeq
$G(-i\tau,\bold{p})$ and $\sigma(\omega,\bold{p})$ are the quantities of major interest for our study. We shall evaluate them in the rest of this paper.\\ 
%In particular from $\chi_M(-i\tau,\bold{x})$
%$\sigma(\omega,\bold{p})$ 
%one can get the z-axis correlator defined, in conformity with lattice calculations, as follows:
Also the z-axis correlator, which, in conformity with lattice calculations, is defined as follows 
\beq
\mathcal{G}(z)\equiv\int\limits_0^\beta d\tau \int d\bold{x_\perp} \chi_M(-i\tau,\bold{x_\perp},z)\;,
\eeq
can be expressed in terms of the finite momentum MSF $\sigma(\omega,\bold{p})$.\\
In fact from Eqs. (\ref{eq:mescorr}) and (\ref{eq:specrit}) it follows that:
\beqa
\mathcal{G}(z) & = & \int\limits_{-\infty}^{+\infty}\frac{dp_z}{2\pi}e^{\displaystyle{i p_z z}}\chi_M(p_0\!=\!0,\bold{p_\perp}\!\!=\!0,p_z)\nonumber\\
{} & = & \int\limits_{-\infty}^{+\infty}\frac{dp_z}{2\pi}e^{\displaystyle{i p_z z}}\int_{-\infty}^{+\infty}d\omega\frac{\sigma(\omega,\bold{p_\perp}\!\!=\!0,p_z)}{\omega}\;.
\eeqa
Clearly, due to rotational invariance, $\chi_M=\chi_M(|\bold{p}|)$ and the choice of the axis is absolutely 
arbitrary: focusing on the z-axis one does not loose any generality.\\
It is justified to conjecture the spatial
 correlations to be exponentially suppressed at large values of $z$, namely:
\beq
\mathcal{G}(z)\underset{z\to +\infty}{\sim}e^{\displaystyle{-m_\trm{scr}z}}\;.
\eeq
Hence from the evaluation of the \emph{meson screening mass} $m_\trm{scr}$, governing the large distance behavior
 of the correlator, one can extract informations on the nature of the excitations characterizing the QGP phase.\\
In the rest of this paper we shall show how the approach employed in Refs.~\cite{berry,mus} can be 
extended to the evaluation of the finite momentum meson spectral functions and temporal correlators. As already anticipated, due to the hard numerical work required to get $\mathcal{G}(z)$, the evaluation of the z-axis correlations is left out for future work.
\section{HTL quark spectral function}\label{sec:quark}
\begin{figure}[!htp]
\begin{center}
\includegraphics[clip,width=0.8\textwidth]{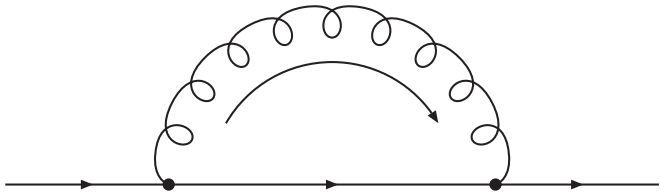}
\caption{The diagram contributing to the quark self-energy in the HTL approximation. The wavy line denotes 
a gluon. The external quark lines are soft.}\label{fig:quarkselfhtl} 
\end{center}
\end{figure}
In this Section, following \cite{bi}, we recall the spectral representation of the HTL resummed quark propagator
 which will be exploited in the evaluation of the mesonic correlators and will result helpful in understanding
 the physical meaning of the different terms contributing to the MSF.\\
We remind the reader that the HTL approximation for the quark propagator amounts to consider the self-energy diagram
 in Fig. \ref{fig:quarkselfhtl}.\\
It can be shown that the major role in dressing the quark propagator is not played by the vacuum fluctuations (a quark emitting and reabsorbing a gluon), but by the interaction with the other particles of the thermal bath (antiquarks and gluons), carrying typical (hard) momenta $K\sim T$.\\
The HTL quark propagator can be cast in the following form \cite{bi}:
\beq\label{delta}
^{\star}S(\omega,\bold{p})=\,^{\star}\Delta_{+}(\omega,p)\frac{\gamma^0-\bold{\gamma\cdot\hat{p}}}{2}
+\,^{\star}\Delta_{-}(\omega,p)\frac{\gamma^0+\bold{\gamma\cdot\hat{p}}}{2}
\eeq
being
\beq\label{den}
^{\star}\Delta_{\pm}(\omega,p)=\frac{-1}{{\displaystyle\omega\mp p-\frac{m_q^2}{2p}\left[\left(1\mp 
\frac{\omega}{p}\right)\ln\frac{\omega+p}{\omega-p}\pm 2\right]}}\; ,
\eeq
with the quark thermal mass $m_q\!=\!g(T)T/\sqrt{6}$, $g(T)$ being the gauge running coupling evaluated at the renormalization scale $\mu\!\sim\!T$.\\
Likewise the following representation
\beq\label{rappr}
^{\star}S(i\omega_n,\bold{p})=-\frac{\gamma^0\!-\!\bold{\gamma\cdot\hat{p}}}{2}\!\int\limits_{-\infty}^{
+\infty}\!d\omega\frac{\rho_{+}(\omega,p)}{i\omega_n-\omega}\, -\, \frac{\gamma^0\!+\!\bold{\gamma\cdot
\hat{p}}}{2}\!\int\limits_{-\infty}^{+\infty}\!d\omega\frac{\rho_{-}(\omega,p)}{i\omega_n-\omega}\; ,
\eeq
is easily proved to hold as well. In (\ref{rappr}) $\rho_{\pm}$ and $^{\star}\Delta_{\pm}$ are related as follows:
\beq\label{defspectral}
^{\star}\Delta_{\pm}(z,p)=-\int\limits_{-\infty}^{+\infty}\!d\omega\frac{\rho_{\pm}(\omega,p)}{z-\omega}\, 
\Rightarrow \rho_{\pm}(\omega,p)=\frac{1}{\pi}\trm{Im}\, ^{\star}\Delta_{\pm}(\omega+i\eta,p)\; ,
\eeq
with $z$ off the real axis.\\
Lumping together the above results one gets the following compact expression for the HTL quark spectral 
function
\beq\label{spectral}
\rho_{\trm{HTL}}(\omega,\bold{p})=\frac{\gamma^0-\bold{\gamma\cdot\hat{p}}}{2}\rho_{+}(\omega,p)\, +\, 
\frac{\gamma^0+\bold{\gamma\cdot\hat{p}}}{2}\rho_{-}(\omega,p) \; ,
\eeq
leading to the spectral representation
\beq\label{fermspec}
^{\star}S(i\omega_{n},\bold{p})=-\int\limits_{-\infty}^{+\infty}\!d\omega\frac{\displaystyle{\rho_{\trm{HTL}}
(\omega,\bold{p})}}{i\omega_n-\omega}\;,
\eeq
for the thermal fermionic Matsubara propagator.\\ 
\begin{figure}[!htp]
\begin{center}
\includegraphics[clip,width=0.7\textwidth]{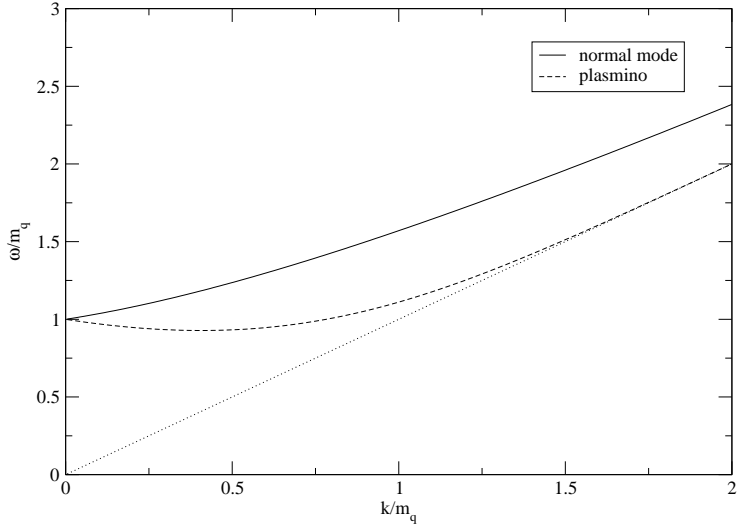}
\caption{Dispersion relations corresponding to the quasiparticle poles of the HTL fermion propagator in the time-like domain. We use the dimensionless units $k/m_q$ and $\omega/m_q$.}\label{fermdisp} 
\end{center}
\end{figure}
\begin{figure}[!htp]
\begin{center}
\includegraphics[clip,width=0.7\textwidth]{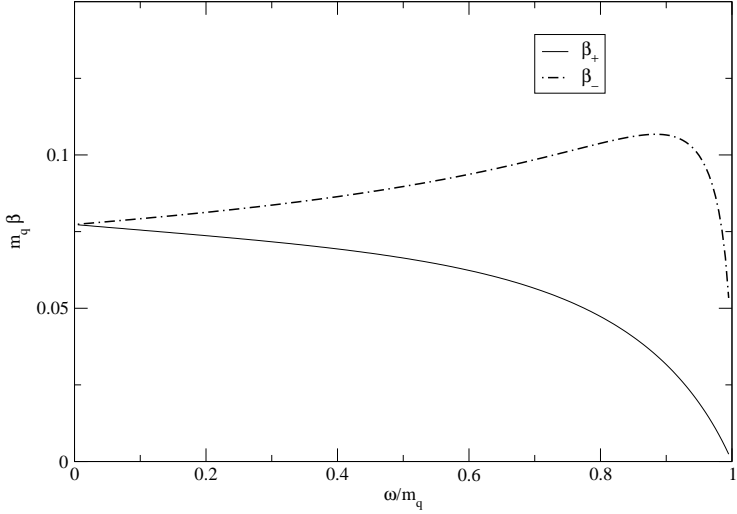}
\caption{Dimensionless spectral function $m_q\rho_\pm$ for space-like momenta at $p=m_q$ as a function
 of $\omega/m_q$. The maximum of $\beta_{-}$ arises from a second solution $\omega_{-}(p)$ of the equation $\trm{Re}(^\star\Delta_-^{-1})=0$  in the space-like region, but in this case, due to the imaginary part 
of the self-energy, it does not give rise to a well defined peak corresponding to a quasi-particle excitation.}
\label{Landau} 
\end{center}
\end{figure}
The explicit expression of the HTL quark spectral function introduced in Eq. (\ref{rappr}) reads \cite{lb}:
\beq
\rho_{\pm}(\omega,k)=\frac{\omega^2-k^2}{2m_q^2}[\delta(\omega-\omega_{\pm})+\delta(\omega+\omega_{\mp})]
+\beta_{\pm}(\omega,k)\theta(k^2-\omega^2)\;,\label{2part}
\eeq
with
\beq
\beta_{\pm}(\omega,k)=-\frac{m_q^2}{2}\frac{\pm\omega-k}{\left[k(-\omega\pm k)+m_q^2\left(\pm1-\frac{
\pm\omega-k}{2k}\ln\frac{k+\omega}{k-\omega}\right)\right]^2+\left[\frac{\pi}{2}m_q^2\frac{\pm\omega-k}
{k}\right]^2}\;,
\eeq
where the two dispersion relations $\omega_\pm(k)$ are shown in Fig. \ref{fermdisp}, while the behavior 
of $\beta_\pm(\omega,k)$ is plotted in Fig. \ref{Landau}. The \emph{thermal gap mass} of the quark $m_q=gT/\sqrt{6}$ shows up in Fig.  \ref{fermdisp} at $k=0$.

The HTL spectral functions in Eq. (\ref{2part}) consist of two pieces: a pole term, arising from the zeros of the real part
 of the denominators of $^{\star}\Delta_{\pm}$ in Eq. (\ref{delta}), and a continuum term, corresponding to
 the \emph{Landau damping} of a quark propagating in the thermal bath.\\
In the time-like domain $\omega>k$ the quark spectral function is characterized  by the quasiparticle poles $\omega\!=\!\omega_\pm(k)$. The dispersion relation $\omega_{+}(k)$ satisfies the equation 
$^{\star}\Delta_{+}^{-1}(\omega_{+}(k),k)=0$ and corresponds to the propagation of a quasi-particle 
with chirality and helicity eigenvalues of the same sign. The dispersion relation  $\omega_{-}(k)$ satisfies instead 
the equation $^{\star}\Delta_{-}^{-1}(\omega_{-}(k),k)=0$ and describes the propagation of an excitation, 
referred to as \textit{plasmino}, with negative helicity over chirality ratio. Note that both these 
excitations are undamped at this level of approximation, since the logarithm in $^{\star}\Delta_{\pm}$
 doesn't develop any imaginary part in the time-like domain.\\
 On the other hand, in the space-like domain $\omega<k$, the quark spectral function gets the continuum contribution $\beta_{\pm}(\omega,k)$ (Landau damping).
Formally such a contribution is related to the imaginary part developed by the logarithm contained in Eq. (\ref{den}).
Physically it reflects two different processes: a quark absorbing a gluon from the thermal bath and being scattered by the latter; a quark annihilating with an antiquark from the thermal bath to give a gluon.\\
In the following we investigate how this structure of the finite temperature quark spectral function affects the finite momentum mesonic correlations.
\section{Mesonic correlations in the pseudoscalar channel}\label{sec:pseudosc}
\begin{figure}[!tp]
\begin{center}
\includegraphics[clip,width=0.3\textwidth]{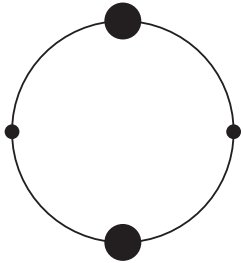}
\caption{Diagram giving the pseudoscalar meson 2-point function in the HTL approximation. The interaction 
vertices are $\gamma^5$ matrices, while the dressed lines are HTL resummed propagators. No vertex correction 
is required within this scheme.}\label{fig:meson2} 
\end{center}
\end{figure}
In the following we shall confine ourselves to the pseudoscalar channel.\\
As shown in \cite{tho}, the pseudoscalar vertex receives no HTL correction . Hence the HTL approximation 
for the meson 2-point function in the pseudoscalar channel, graphically displayed in Fig. \ref{fig:meson2},
 amounts to simply employ for the fermionic lines the resummed propagators given by Eq. (\ref{delta}). 
This leads to the following expression:
\beq
\chi^{\trm{ps}}(i\omega_l,\bold{p})=2N_c\!\frac{1}{\beta}\!\sum_{n=-\infty}^{+\infty}\!\!\int\!\frac{d^3k}
{(2\pi)^3}\trm{Tr}[\gamma^{5}\, ^{\star}S(i\omega_n,\bold{k})\gamma^{5}\, ^{\star}S(i\omega_n\!-i\omega_l,
\bold{k}-\bold{p})]\;.
\eeq
Setting $\bold{q}=\bold{k}-\bold{p}$ and making use of the spectral representation (\ref{fermspec}) of the 
quark propagator one gets:
\begin{multline}
\chi^{\trm{ps}}(i\omega_l,\bold{p})\!=\!2N_c\frac{1}{\beta}\!\sum_{n=-\infty}^{+\infty}\!\!\int\!\frac{d^3k}
{(2\pi)^3}\int\limits_{-\infty}^{+\infty}d\omega_1\int\limits_{-\infty}^{+\infty}d\omega_2\frac{1}{i\omega_n-\omega_1}
\frac{1}{i\omega_n-i\omega_l-\omega_2}\times\\
\times\trm{Tr}[\gamma^5\rho_{\trm{HTL}}(\omega_1,\bold{k})\gamma^5\rho_{\trm{HTL}}(\omega_2,\bold{q})]
\;.\label{eq:mesonHTL}
\end{multline}
The above allows us to extend the work done in Ref.~\cite{berry} for the MSF and the temporal correlator 
to the case of finite spatial momentum. 
\subsection{Finite momentum spectral function: analytical expressions}\label{sec:spec}
%\subsubsection{Analytical expressions}\label{spec_an}
Summing over the Matsubara frequencies in Eq. (\ref{eq:mesonHTL}) with a standard contour integration, 
performing the usual analytical continuation $i\omega_{l}\rightarrow\omega +i\eta^+$ (retarded boundary 
conditions) and taking the imaginary part of the result thus obtained one gets for the MSF the expression:
\begin{multline}
\sigma^{\trm{ps}}(\omega,\bold{p})=-2N_c\int\!\frac{d^3k}{(2\pi)^3}(e^{\beta\omega}-1)
\int\limits_{-\infty}^{+\infty}d\omega_1\int\limits_{-\infty}^{+\infty}d\omega_2\tilde{n}(\omega_1)\tilde{n}(\omega_2)\times\\
\times\delta(\omega-\omega_1-\omega_2)\cdot\trm{Tr}[\gamma^5\rho_{\trm{HTL}}(\omega_1,\bold{k})
\gamma^5\rho_{\trm{HTL}}(-\omega_2,\bold{q})]\;.\label{eq:sigmap}
\end{multline}
Now, inserting Eq. (\ref{spectral}) into Eq. (\ref{eq:sigmap}) and since
\bseq
\begin{align}
\trm{Tr}\left[\gamma^5\frac{\gamma^0\mp\bold{\gamma\cdot\hat{k}}}{2}\gamma^5\frac{\gamma^0\mp
\bold{\gamma\cdot\hat{q}}}{2}\right]&=-(1-\bold{\hat{k}\cdot\hat{q}}),\\
\trm{Tr}\left[\gamma^5\frac{\gamma^0\mp\bold{\gamma\cdot\hat{k}}}{2}\gamma^5\frac{\gamma^0\pm
\bold{\gamma\cdot\hat{q}}}{2}\right]&=-(1+\bold{\hat{k}\cdot\hat{q}})\;,
\end{align}
\eseq
one gets \cite{mus}
\begin{multline}
\sigma^{\trm{ps}}(\omega,\bold{p})\!=\!2N_c\int\!\frac{d^3k}{(2\pi)^3}(e^{\beta\omega}-1)
\int\limits_{-\infty}^{+\infty}d\omega_1\int\limits_{-\infty}^{+\infty}d\omega_2\tilde{n}(\omega_1)\tilde{n}
(\omega_2)\delta(\omega-\!\omega_1-\!\omega_2)\times\\
\times\left\{(1+\bold{\hat{k}\cdot\hat{q}})[\rho_+(\omega_1,k)\rho_-(-\omega_2,q)+\rho_-(\omega_1,k)
\rho_+(-\omega_2,q)]+\right.\\
+\left.(1-\bold{\hat{k}\cdot\hat{q}})[\rho_+(\omega_1,k)\rho_+(-\omega_2,q)+\rho_-(\omega_1,k)
\rho_-(-\omega_2,q)]\right\}\;.\label{eq:sigma0}
\end{multline}
The above expression can also be written as
\begin{multline}
\sigma^{\trm{ps}}(\omega,\bold{p})\!=\!2N_c\int\!\frac{d^3k}{(2\pi)^3}(e^{\beta\omega}-1)
\int\limits_{-\infty}^{+\infty}d\omega_1\int\limits_{-\infty}^{+\infty}d\omega_2\tilde{n}(\omega_1)\tilde{n}
(\omega_2)\delta(\omega-\!\omega_1-\!\omega_2)\times\\
\times\left\{(1+\bold{\hat{k}\cdot\hat{q}})[\rho_+(\omega_1,k)\rho_+(\omega_2,q)+\rho_-(\omega_1,k)
\rho_-(\omega_2,q)]+\right.\\
+\left.(1-\bold{\hat{k}\cdot\hat{q}})[\rho_+(\omega_1,k)\rho_-(\omega_2,q)+\rho_-(\omega_1,k)
\rho_+(\omega_2,q)]\right\}\;.\label{eq:sigmapcompl}
\end{multline}
where the identity $\rho_{+}(-\omega,k)=\rho_{-}(\omega,k)$ has been used.\\
As discussed in Sec. \ref{sec:quark} the HTL quark spectral function $\rho_{\pm}(\omega,k)$, whose expression in given in Eq. (\ref{2part}), reflects the singularities of the quark propagator (\ref{den}) in the complex $\omega$-plane. These lie on the real $\omega$-axis. At a given value of the spatial momentum $k$, in the time-like domain ($\omega^2>k^2$) discrete poles are associated to quasiparticle excitations; a cut for space-like momenta ($\omega^2<k^2$) is related to the Landau damping.\\ 
Inserting the explicit expressions for $\rho_{\pm}(\omega,k)$ into Eq. (\ref{eq:sigma0}) one gets then three different contributions (pole-pole, pole-cut and cut-cut) to the HTL pseudoscalar spectral MSF, namely:
\beq
\sigma^{\trm{ps}}_{\trm{HTL}}(\omega,\bold{p})=\sigma^{\trm{pp}}(\omega,\bold{p})+\sigma^{\trm{pc}}
(\omega,\bold{p})+\sigma^{\trm{cc}}(\omega,\bold{p})\;.
\eeq
Below we give the explicit expressions for these three terms.\\
In the following
\beq
Z_{\pm}(k)=\frac{\omega_{\pm}^2(k) - k^2}{2 m_q^2}
\eeq
are the residues of the quasi-particle poles.\\
Moreover we shall denote with $q_+$ ($\bar{q}_+$) the normal quark (antiquark) mode, with $q_-$ ($\bar{q}_-$) the plasmino (antiplasmino) mode and with $M$ the excitation carrying the quantum numbers of a pseudoscalar meson.

The pole-pole contribution (for $\omega>0$) is made up of twelve terms and reads:
\begin{multline}
\sigma^{\trm{pp}}(\omega,p)=\frac{N_c}{2\pi^2} (e^{\beta\omega}-1) \int\limits^{+1}_{-1} dx \\
\left\{ \sum_{k_1} (1+{\hat{\bold{k}}_1\cdot\hat{\bold{q}}_1})\tilde{n}(\omega_+(k_1))\tilde{n}(\omega_+(q_1))
Z_+(k_1) Z_+(q_1)\frac{k_1^2}{|\omega'_+(k_1)+\omega'_+(q_1) |}+\right.\\
+\sum_{k_2} (1+{\hat{\bold{k}}_2\cdot\hat{\bold{q}}_2})\tilde{n}(\omega_+(k_2))[1-\tilde{n}(\omega_-(q_2))]
 Z_+(k_2) Z_-(q_2)\frac{k_2^2}{|\omega'_+(k_2)-\omega'_-(q_2)|}\\
 +\sum_{k_3} (1+{\hat{\bold{k}}_3\cdot\hat{\bold{q}}_3})[1-\tilde{n}(\omega_-(k_3))]\tilde{n}(\omega_+(q_3))
Z_-(k_3) Z_+(q_3)\frac{k_3^2}{|\omega'_-(k_3)-\omega'_+(q_3) |}\\
+\sum_{k_4} (1+{\hat{\bold{k}}_4\cdot\hat{\bold{q}}_4})\tilde{n}(\omega_-(k_4))\tilde{n}(\omega_-(q_4))
 Z_-(k_4) Z_-(q_4)\frac{k_4^2}{|\omega'_-(k_4)+\omega'_-(q_4)|}\\
+\sum_{k_5} (1+{\hat{\bold{k}}_5\cdot\hat{\bold{q}}_5})\tilde{n}(\omega_-(k_5))[1-\tilde{n}(\omega_+(q_5))]
 Z_-(k_5) Z_+(q_5)\frac{k_5^2}{|\omega'_-(k_5)-\omega'_+(q_5)|}\\
 +\sum_{k_6} (1+{\hat{\bold{k}}_6\cdot\hat{\bold{q}}_6})[1-\tilde{n}(\omega_+(k_6))]\tilde{n}(\omega_-(q_6))
Z_+(k_6) Z_-(q_6)\frac{k_6^2}{|\omega'_+(k_6)-\omega'_-(q_6) |}\\ 
+\sum_{k_7} (1-{\hat{\bold{k}}_7\cdot\hat{\bold{q}}_7})\tilde{n}(\omega_+(k_7))\tilde{n}(\omega_-(q_7))
Z_+(k_7) Z_-(q_7)\frac{k_7^2}{|\omega'_+(k_7)+\omega'_-(q_7) |}\\
+\sum_{k_8} (1-{\hat{\bold{k}}_8\cdot\hat{\bold{q}}_8})\tilde{n}(\omega_+(k_8))[1-\tilde{n}(\omega_+(q_8))]
 Z_+(k_8) Z_+(q_8)\frac{k_8^2}{|\omega'_+(k_8)-\omega'_+(q_8)|}\\
 +\sum_{k_9} (1-{\hat{\bold{k}}_9\cdot\hat{\bold{q}}_9})[1-\tilde{n}(\omega_-(k_9))]\tilde{n}(\omega_-(q_9))
Z_-(k_9) Z_-(q_9)\frac{k_9^2}{|\omega'_-(k_9)-\omega'_-(q_9) |}\\
+\sum_{k_{10}} (1-{\hat{\bold{k}}_{10}\cdot\hat{\bold{q}}_{10}})\tilde{n}(\omega_-(k_{10}))
\tilde{n}(\omega_+(q_{10}))
 Z_-(k_{10}) Z_+(q_{10})\frac{k_{10}^2}{|\omega'_-(k_{10})+\omega'_+(q_{10})|}+\\
\sum_{k_{11}} (1-{\hat{\bold{k}}_{11}\cdot\hat{\bold{q}}_{11}})\tilde{n}(\omega_-(k_{11}))
[1-\tilde{n}(\omega_-(q_{11}))]
 Z_-(k_{11}) Z_-(q_{11})\frac{k_{11}^2}{|\omega'_-(k_{11})-\omega'_-(q_{11})|}+\\
\left. \sum_{k_{12}} (1-{\hat{\bold{k}}_{12}\cdot\hat{\bold{q}}_{12}})[1-\tilde{n}(
\omega_+(k_{12}))]\tilde{n}(\omega_+(q_{12}))
Z_+(k_{12}) Z_+(q_{12})\frac{k_{12}^2}{|\omega'_+(k_{12})-\omega'_+(q_{12}) |} \right\}.\\\label{eq:sigmapp} 
\end{multline}
In the above the $q_n=\sqrt{k_n^2+p^2-2pk_n x}$ and the $k_n$ are then the solutions of the following equations, corresponding to the physical processes explicitly specified below:
\beqa
1) & \omega-\omega_+(k_1)-\omega_+(q_1)=0 &
 \qquad q_+ +\bar{q}_+\rightarrow M \nonumber \\
2) & \omega-\omega_+(k_2)+\omega_-(q_2)=0 &
 \qquad q_+\rightarrow q_- + M\nonumber \\
3) & \omega+\omega_-(k_3)-\omega_+(q_3)=0 &
 \qquad \bar{q}_+ \rightarrow \bar{q}_- + M\nonumber \\
4) & \omega-\omega_-(k_4)-\omega_-(q_4)=0 &
 \qquad q_- +\bar{q}_-\rightarrow M \nonumber \\
5) & \omega-\omega_-(k_5)+\omega_+(q_5)=0 &
 \qquad q_-\rightarrow q_+ + M\nonumber \\
6) & \omega+\omega_+(k_6)-\omega_-(q_6)=0 &
 \qquad \bar{q}_- \rightarrow \bar{q}_+ + M\nonumber \\
7) & \omega-\omega_+(k_7)-\omega_-(q_7)=0 &
 \qquad q_+ +\bar{q}_-\rightarrow M\nonumber \\
8) & \omega-\omega_+(k_8)+\omega_+(q_8)=0 &
 \qquad q_+\rightarrow q_+ + M\nonumber \\
9) & \omega+\omega_-(k_9)-\omega_-(q_9)=0 &
 \qquad \bar{q}_- \rightarrow \bar{q}_- + M\nonumber \\
10) & \omega-\omega_-(k_{10})-\omega_+(q_{10})=0 &
 \qquad q_- +\bar{q}_+\rightarrow M\nonumber \\
11) & \omega-\omega_-(k_{11})+\omega_-(q_{11})=0 &
 \qquad q_-\rightarrow q_- + M\nonumber \\
12) & \omega+\omega_+(k_{12})-\omega_+(q_{12})=0 &
 \qquad \bar{q}_+ \rightarrow \bar{q}_+ + M\,.\label{eq:processes}
\eeqa
The above identifications follow quite easily after inserting into Eq. (\ref{eq:sigma0}) the quasiparticle contributions of the quark spectral functions given by the equations
\bseq
\begin{align}
\rho_+^\trm{pole}(\omega,k)&=\underbrace{Z_+(k)\delta(\omega-\omega_+(k))}_{\trm{quark mode}}+\underbrace{Z_-(k)\delta(\omega+\omega_-(k))}_{\trm{antiplasmino}}\\
\rho_-^\trm{pole}(\omega,k)&=\underbrace{Z_-(k)\delta(\omega-\omega_-(k))}_{\trm{plasmino}}+\underbrace{Z_+(k)\delta(\omega+\omega_+(k))}_{\trm{antiquark mode}}\;.
\end{align}
\eseq
The role played by the statistical distributions in Eq. (\ref{eq:sigmapp}) appears then clear. In the annihilation processes the initial particles belong to occupied states (hence the product of two Fermi distributions). In the decay processes the initial particle belongs to an occupied state (hence the Fermi distribution), while the final one is produced in an empty state (hence the Pauli blocking factor).\\
The pole-cut contribution is found to consist of sixteen terms and reads:
\begin{multline}
\sigma^{\trm{pc}}(\omega,p)  =  \frac{N_c}{2\pi^2} (e^{\beta\omega}-1)
\int\limits_{0}^{\infty}dk\,k^2   \int\limits_{-1}^{+1}d\cos\theta \left\{ (1+\bold{\hat{k}\cdot\hat{q}})\times\right.\\
\times \left[Z_+(k)\beta_{+}(\omega-\omega_+(k),q) \tilde{n}(\omega_+(k)) \tilde{n}(\omega-\omega_+(k))
 \theta(q^2-(\omega-\omega_{+}(k))^2)\right. + \\
+ Z_-(k)\beta_{+}(\omega+\omega_-(k),q) (1-\tilde{n}(\omega_-(k))) \tilde{n}(\omega+\omega_-(k))
 \theta(q^2-(\omega+\omega_{-}(k))^2) + \\
+ Z_+(q)\beta_{+}(\omega-\omega_+(q),k) \tilde{n}(\omega_+(q)) \tilde{n}(\omega-\omega_+(q))
 \theta(k^2-(\omega-\omega_{+}(q))^2) + \\
+ Z_-(q)\beta_{+}(\omega+\omega_-(q),k) (1-\tilde{n}(\omega_-(q))) \tilde{n}(\omega+\omega_-(q))
 \theta(k^2-(\omega+\omega_{-}(q))^2) + \\
+ Z_-(k)\beta_{-}(\omega-\omega_-(k),q) \tilde{n}(\omega_-(k)) \tilde{n}(\omega-\omega_-(k))
 \theta(q^2-(\omega-\omega_{-}(k))^2) + \\
+ Z_+(k)\beta_{-}(\omega+\omega_+(k),q) (1-\tilde{n}(\omega_+(k))) \tilde{n}(\omega+\omega_+(k))
 \theta(q^2-(\omega+\omega_{+}(k))^2) + \\
Z_-(q)\beta_{-}(\omega-\omega_-(q),k) \tilde{n}(\omega_-(q)) \tilde{n}(\omega-\omega_-(q))
 \theta(k^2-(\omega-\omega_{-}(q))^2) + \\
\left. +Z_+(q)\beta_{-}(\omega+\omega_+(q),k) (1-\tilde{n}(\omega_+(q))) \tilde{n}(\omega+\omega_+(q))
 \theta(k^2-(\omega+\omega_{+}(q))^2)\right]+\\
+ (1-\bold{\hat{k}\cdot\hat{q}}) \times \\
 \times \left[Z_+(k)\beta_{-}(\omega-\omega_+(k),q) \tilde{n}(\omega_+(k)) \tilde{n}(\omega-\omega_+(k))
 \theta(q^2-(\omega-\omega_{+}(k))^2)\right. + \\
+ Z_-(k)\beta_{-}(\omega+\omega_-(k),q) (1-\tilde{n}(\omega_-(k))) \tilde{n}(\omega+\omega_-(k))
 \theta(q^2-(\omega+\omega_{-}(k))^2) + \\
+ Z_-(q)\beta_{+}(\omega-\omega_-(q),k) \tilde{n}(\omega_-(q)) \tilde{n}(\omega-\omega_-(q))
 \theta(k^2-(\omega-\omega_{-}(q))^2) + \\
+ Z_+(q)\beta_{+}(\omega+\omega_+(q),k) (1-\tilde{n}(\omega_+(q))) \tilde{n}(\omega+\omega_+(q))
 \theta(k^2-(\omega+\omega_{+}(q))^2) + \\
+ Z_-(k)\beta_{+}(\omega-\omega_-(k),q) \tilde{n}(\omega_-(k)) \tilde{n}(\omega-\omega_-(k))
 \theta(q^2-(\omega-\omega_{-}(k))^2) + \\
+ Z_+(k)\beta_{+}(\omega+\omega_+(k),q) (1-\tilde{n}(\omega_+(k))) \tilde{n}(\omega+\omega_+(k))
 \theta(q^2-(\omega+\omega_{+}(k))^2) + \\
Z_+(q)\beta_{-}(\omega-\omega_+(q),k) \tilde{n}(\omega_+(q)) \tilde{n}(\omega-\omega_+(q))
 \theta(k^2-(\omega-\omega_{+}(q))^2) + \\
\left.\left. + Z_-(q)\beta_{-}(\omega+\omega_-(q),k) (1-\tilde{n}(\omega_-(q))) \tilde{n}(\omega+\omega_-(q))
 \theta(k^2-(\omega+\omega_{-}(q))^2)\right]\right\},\\
\end{multline}
where the trivial azimuthal integration has been performed and only the integration over the polar angle $\theta$ between $\bold{\hat{k}}$ and $\bold{\hat{p}}$ is left out.
Finally the cut-cut contribution is given by:
\begin{multline}
\sigma^{\trm{cc}}(\omega,p)=\frac{N_c}{2 \pi^2} \int\limits_{-1}^{+1} d\cos\theta \int\limits_0^\infty dk\,k^2\int\limits_{-k}^{+k}d\omega_1 \,\tilde{n}(\omega_1)\tilde{n}(\omega-\omega_1)
\theta(q^2-(\omega-\omega_1)^2)\\
  (e^{\beta\omega}-1)\left\{(1+\bold{\hat{k}\cdot\hat{q}}) \left[
\beta_+(\omega_1,k)\beta_+(\omega-\omega_1,q)+\beta_-(\omega_1,k)\beta_-(\omega-\omega_1,q)\right]\right. +\\
\left. +(1-\bold{\hat{k}\cdot\hat{q}}) \left[
\beta_+(\omega_1,k)\beta_-(\omega-\omega_1,q)+\beta_-(\omega_1,k)\beta_+(\omega-\omega_1,q)\right]\right\}.\\
\end{multline}
For the sake of comparison we give below the expression of the free MSF at 
finite momentum, reported also in \cite{aar}, which can be obtained, 
for massive quarks, inserting into Eq. (\ref{eq:sigmap})
the free quark spectral function 
\beq
\rho_F(K)=(K\hspace{-.25cm}{\slash}+m)\frac{1}{2\epsilon_k}
[\delta(k_0-\epsilon_k)-\delta(k_0+\epsilon_k)]\,.
\eeq
One obtains:
\begin{multline}
\sigma^{\trm{ps}}_{\trm{free}}(\omega,\bold{p})\!=\!N_c N_f
\int\!\frac{d^3k}{(2\pi)^3}\times\\
\times\left\{ \left(1\!+\!\frac{\bold{k\cdot q}}{\epsilon_k\epsilon_q}\!
+\!\frac{m^2}{\epsilon_k\epsilon_q}\right)(1\!-\!\tilde{n}(\epsilon_k)\!
-\!\tilde{n}(\epsilon_q))[\delta(\omega\!-\!(\epsilon_k\!+\!\epsilon_q))
-\delta(\omega\!+\!(\epsilon_k\!+\!\epsilon_q))]+\right.\\
+\left.\left(1\!-\!\frac{\bold{k\cdot q}}{\epsilon_k\epsilon_q}
\!-\!\frac{m^2}{\epsilon_k\epsilon_q}\right)(\tilde{n}(\epsilon_q)
\!-\!\tilde{n}(\epsilon_k))[\delta(\omega\!-\!(\epsilon_k\!-\!\epsilon_q))
-\delta(\omega\!+\!(\epsilon_k\!-\!\epsilon_q))]\right\}\;,
\end{multline}
where $\epsilon_k = \sqrt{k^2+m^2}$ and $K=(k_0,\bold{k})$.\\
The delta functions can be exploited to perform the integration over the momentum $\bold{k}$.
The general formula for the free spectral function for an arbitrary quark mass $m$ and number of flavors $N_f$ is then:
\begin{multline}\label{eq:freefinite}
\sigma^{\trm{ps}}_\trm{free}(\omega,\bold{p})\!=\! \frac{N_c N_f}{8
\pi^2} (\omega^2 - p^2)\times\\
\times\left\{\theta(\omega^2-p^2-4 m^2)\left[\sqrt{1-\frac{4 m^2}{\omega^2-p^2}}+\frac{2}{p \beta}A\right]+\theta(p^2-\omega^2)\frac{2}{p \beta} B \right\}\;,
\end{multline}
where $\beta=1/T$ and
\beqa
A = \log\left(1+e^{-\frac{\beta}{2}\Big(\omega+p\sqrt{1-\frac{4 m^2}{\omega^2-p^2}}\Big)}\right)
\!-\!\log\left(1+e^{-\frac{\beta}{2}\Big(\omega-p\sqrt{1-\frac{4 m^2}{\omega^2-p^2}}\Big)}\right)\,,
\\
B = \log\left(1+e^{-\frac{\beta}{2}\Big(\omega+p\sqrt{1-\frac{4 m^2}{\omega^2-p^2}}\Big)}\right)\!
-\!\log\left(1+e^{+\frac{\beta}{2}\Big(\omega-p\sqrt{1-\frac{4 m^2}{\omega^2-p^2}}\Big)}\right) \,.
\eeqa
For $m=0$ we have:
\beq
\sigma^{\trm{ps}}_\trm{free}(\omega,\bold{p})\!=\! \frac{N_c N_f}{8
\pi^2} (\omega^2\!-\!p^2) \Big\{\theta(\omega^2\!-\!p^2)\Big[1\!+\!\frac{2}{p \beta} A \Big] + \theta(p^2\!-\!\omega^2) \frac{2}{p \beta} B \Big\}.
\eeq
In the limit where $ p \rightarrow 0$ Eq. (\ref{eq:freefinite}) becomes:
\beq
\sigma^{\trm{ps}}_\trm{free}(\omega,0)\!=\! \frac{N_c N_f}{8\pi^2} 
\theta(\omega^2 - 4 m^2) \omega^2 \sqrt{1-\frac{4 m^2}{\omega^2}} 
\Big(1- \frac{2}{e^{\frac{\beta \omega}{2}}+1}\Big)\;,
\eeq
in agreement with what quoted in \cite{berry,mus}.
\subsection{Finite momentum spectral function: numerical results}\label{spec_num}
In this Section we present our numerical results for the finite momentum HTL MSF in the pseudoscalar channel. We remind the reader that the HTL approximation has been shown to provide good results for different thermodynamical observables (namely in agreement with the ones given by lattice calculations) starting from $T\approx2.5T_c$ \cite{bie,blasu,blater}. Nevertheless in the following, for the sake of completeness, we will often present results also referring to lower temperatures.

\begin{figure}[!htp]
\begin{center}
\includegraphics[clip,width=0.75\textwidth]{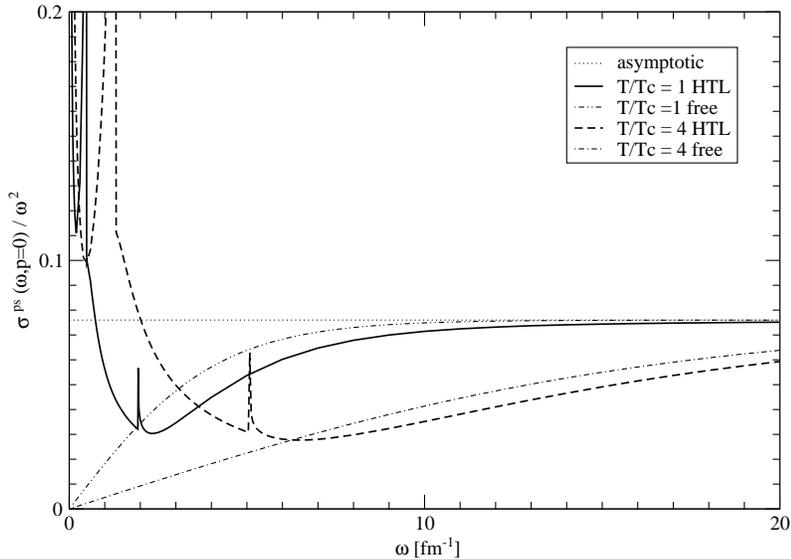}
\caption{The zero momentum HTL and free MSF (divided by $\omega^2$) as a function of $\omega$ for two different temperatures. Also shown is the asymptotic high energy plateau.}
\label{Fig:spec_p0} 
\end{center}
\end{figure}
In order to have a reference benchmark, we start by considering the zero momentum case. In Fig. \ref{Fig:spec_p0} we plot the spectral function divided by $\omega^2$, in order to get rid of the uninteresting asymptotic high energy growth. Two different temperatures are considered. One can appreciate the dramatic difference in the behavior at low energy between the free curves (which vanish at $\omega=0$) and the HTL ones (which diverge for $\omega\to0$). One can also recognize the Van Hove (V.H.) singularities in the HTL curves arising from a divergence in the density of states due to the minimum in the plasmino dispersion relation. All the curves approach for large values of $\omega$ the same high energy, temperature independent, plateau.

\begin{figure}[!htp]
\begin{center}
\includegraphics[clip,width=0.75\textwidth]{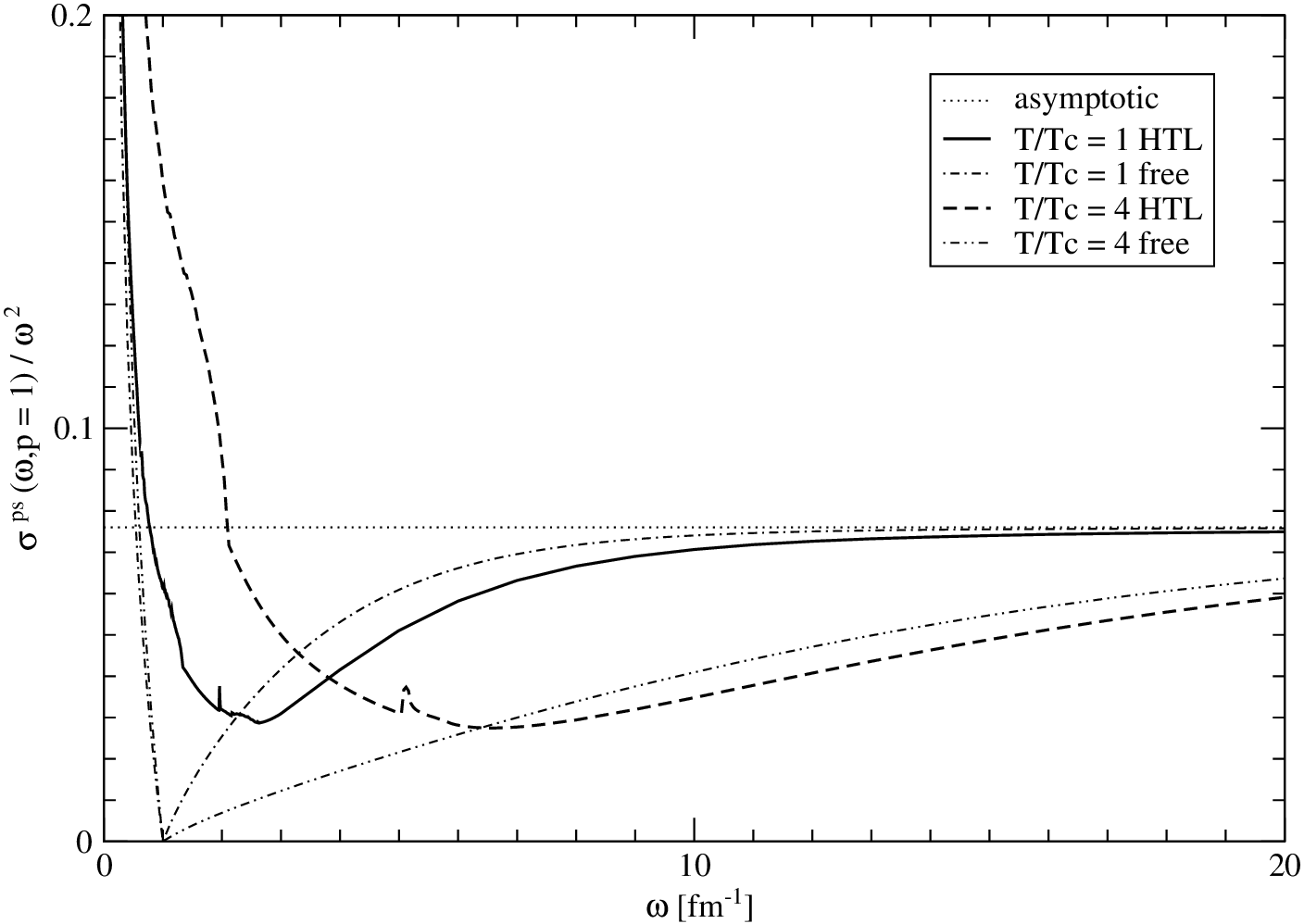}
\caption{The finite momentum HTL and free MSF (divided by $\omega^2$) as a function of $\omega$ at $p=1$ fm$^{-1}$ for two different temperatures. Also shown is the asymptotic high energy plateau.}
\label{Fig:spec_w2p1} 
\end{center}
\end{figure}
\begin{figure}[!htp]
\begin{center}
\includegraphics[clip,width=0.75\textwidth]{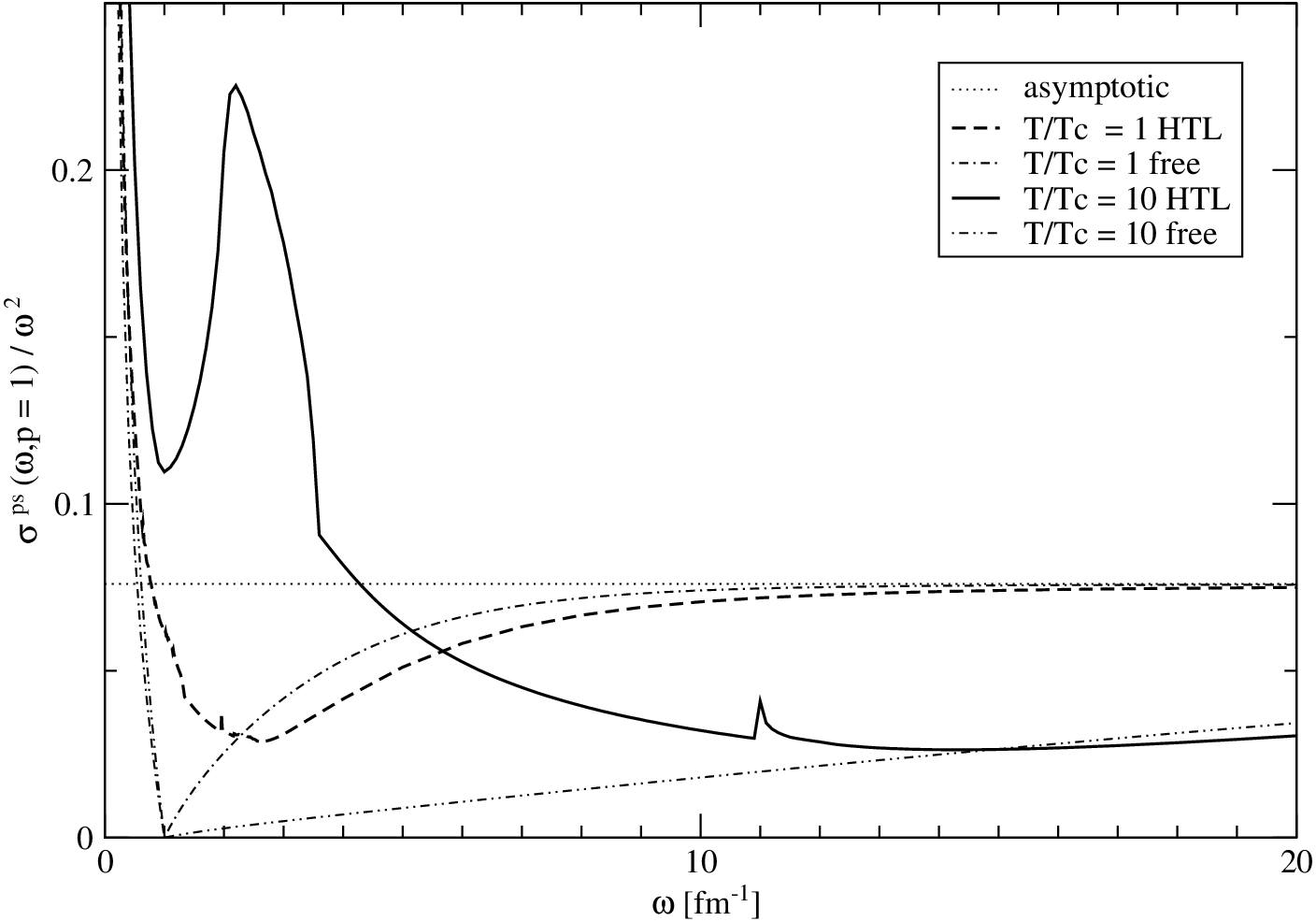}
\caption{The finite momentum HTL and free MSF (divided by $\omega^2$) as a function of $\omega$ at $p=1$ fm$^{-1}$ for two different temperatures. Also shown is the asymptotic high energy plateau.}
\label{Fig:spec_w2p1new} 
\end{center}
\end{figure}
We now investigate how things change at finite spatial momentum. In Figs. \ref{Fig:spec_w2p1} and \ref{Fig:spec_w2p1new} we plot the pseudoscalar MSF (divided by $\omega^2$) for $p=1$ fm$^{-1}$ at different temperatures. The following general features concern all the considered cases: the (massless) non-interacting result vanishes on the light-cone (i.e. for $\omega=p$), at variance with the HTL curves, which stays finite there. On the other hand when $\bold{p}$ is finite both the free and the interacting curves diverge as $\omega\to0$. Furthermore in the HTL case the V.H. singularities are washed-out by the angular integration in Eq. (\ref{eq:sigmapp}). Finally both the free and the HTL finite momentum MSF approach the asymptotic plateau for large values of $\omega$. 

\begin{figure}[!htp]
\begin{center}
\includegraphics[clip,width=0.75\textwidth]{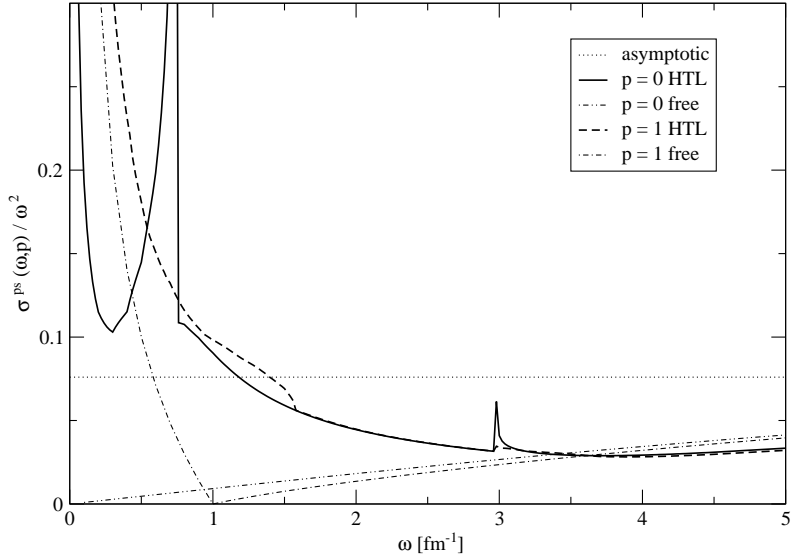}
\caption{The HTL and free MSF (divided by $\omega^2$) as a function of $\omega$ at $p=0$ and $p=1$ fm$^{-1}$ for $T=2T_c$. Also shown is the asymptotic high energy plateau.}
\label{Fig:spec_p01} 
\end{center}
\end{figure}
\begin{figure}[!htp]
\begin{center}
\includegraphics[clip,width=0.75\textwidth]{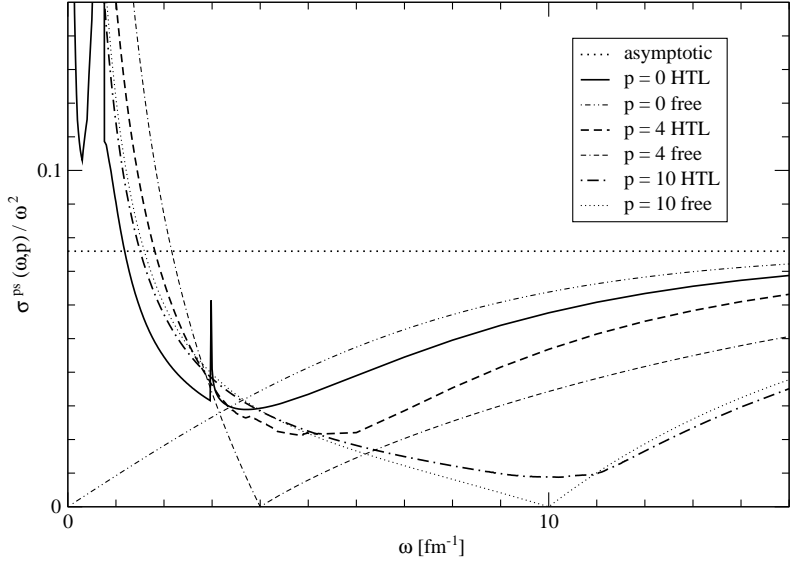}
\caption{The HTL and free MSF (divided by $\omega^2$) as a function of $\omega$ at $T=2T_c$ for different values of the spatial momentum (in fm$^{-1}$). Also shown is the asymptotic high energy plateau.}
\label{Fig:spec_p010} 
\end{center}
\end{figure}
In Figs. \ref{Fig:spec_p01} and \ref{Fig:spec_p010} we display the MSF (divided by $\omega^2$) at $T=2T_c$ for a range of momenta from $p=0$ to $p=10$ fm$^{-1}$. We can thus follow its momentum evolution.

\begin{figure}[!htp]
\begin{center}
\includegraphics[clip,width=0.65\textwidth]{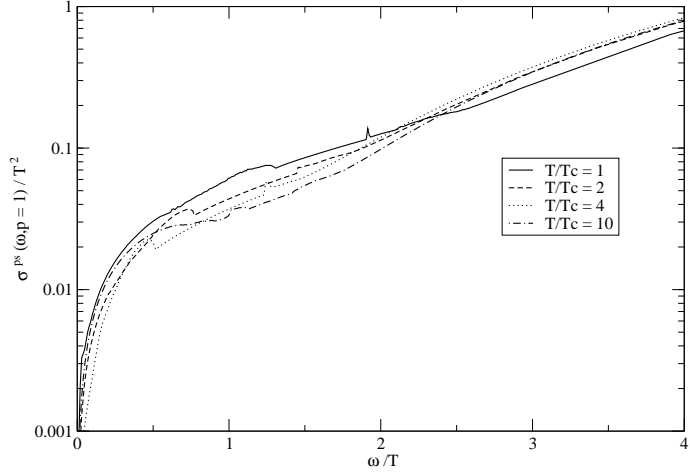}
\caption{The finite momentum HTL MSF (in units of $T^2$) as a function of $\omega/T$ for $p=1$ fm$^{-1}$ at different temperatures.}
\label{Fig:spec_p1} 
\end{center}
\end{figure}
\begin{figure}[!htp]
\begin{center}
\includegraphics[clip,width=0.65\textwidth]{p4.eps}
\caption{The same as in Fig. \ref{Fig:spec_p1} for $p=4$ fm$^{-1}$.}
\label{Fig:spec_p4} 
\end{center}
\end{figure}
\begin{figure}[!htp]
\begin{center}
\includegraphics[clip,width=0.65\textwidth]{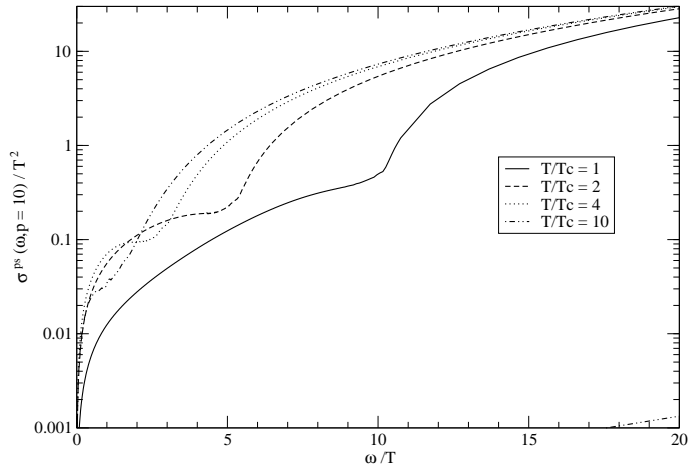}
\caption{The same as in Fig. \ref{Fig:spec_p1} for $p=10$ fm$^{-1}$.}
\label{Fig:spec_p10} 
\end{center}
\end{figure}
\begin{figure}[!htp]
\begin{center}
\includegraphics[clip,width=0.65\textwidth]{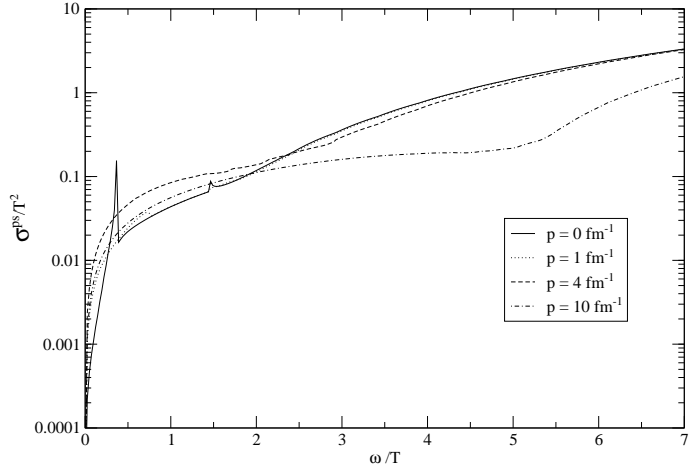}
\caption{The HTL MSF (in units of $T^2$) as a function of $\omega/T$ for different values of the spatial momentum. The plot refers to $T=2T_c$.}
\label{Fig:spec2tc} 
\end{center}
\end{figure}
\begin{figure}[!htp]
\begin{center}
\includegraphics[clip,width=0.65\textwidth]{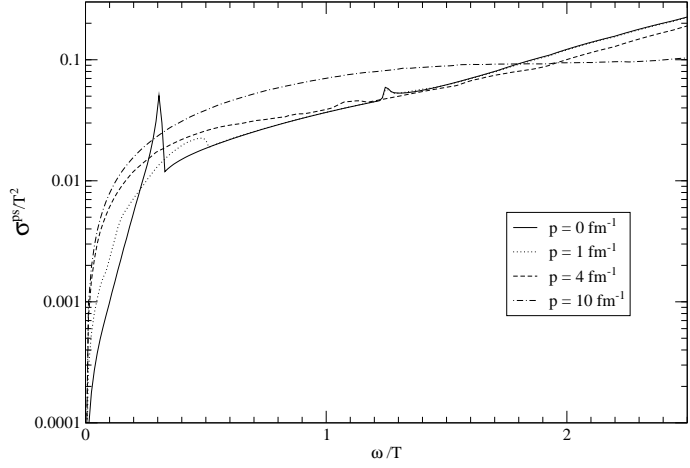}
\caption{The same as in Fig. \ref{Fig:spec2tc} for $T=4T_c$.}
\label{Fig:spec4tc} 
\end{center}
\end{figure}
\begin{figure}[!htp]
\begin{center}
\includegraphics[clip,width=0.65\textwidth]{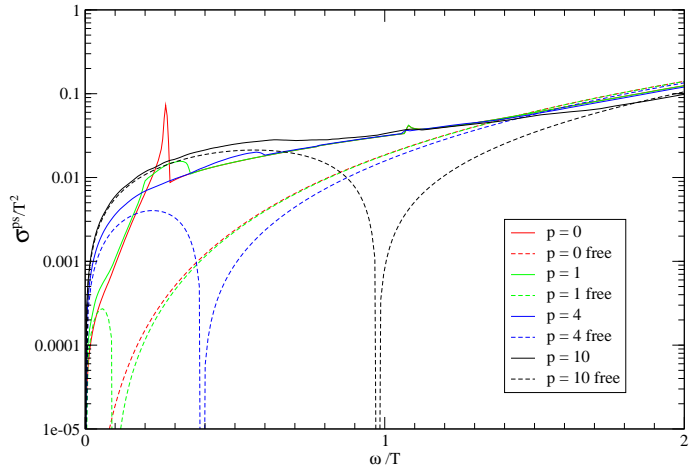}
\caption{The same as in Fig. \ref{Fig:spec2tc} for $T=10T_c$. Also shown are the (massless) non-interacting curves, which vanish on the light-cone.}
\label{Fig:spec10tc} 
\end{center}
\end{figure}
In Figs. \ref{Fig:spec_p1}, \ref{Fig:spec_p4} and \ref{Fig:spec_p10} we display, for a range of temperatures from $T=T_c$ to $T=10T_c$, the finite momentum HTL MSF for $p=1,4,10$  fm$^{-1}$ respectively.

Furthermore in Figs. \ref{Fig:spec2tc}, \ref{Fig:spec4tc} and \ref{Fig:spec10tc} we display how, at a given temperature, an increasing value of the spatial momentum affects the behavior of the MSF.

\begin{figure}[!htp]
\begin{center}
\includegraphics[clip,width=0.70\textwidth,angle=270]{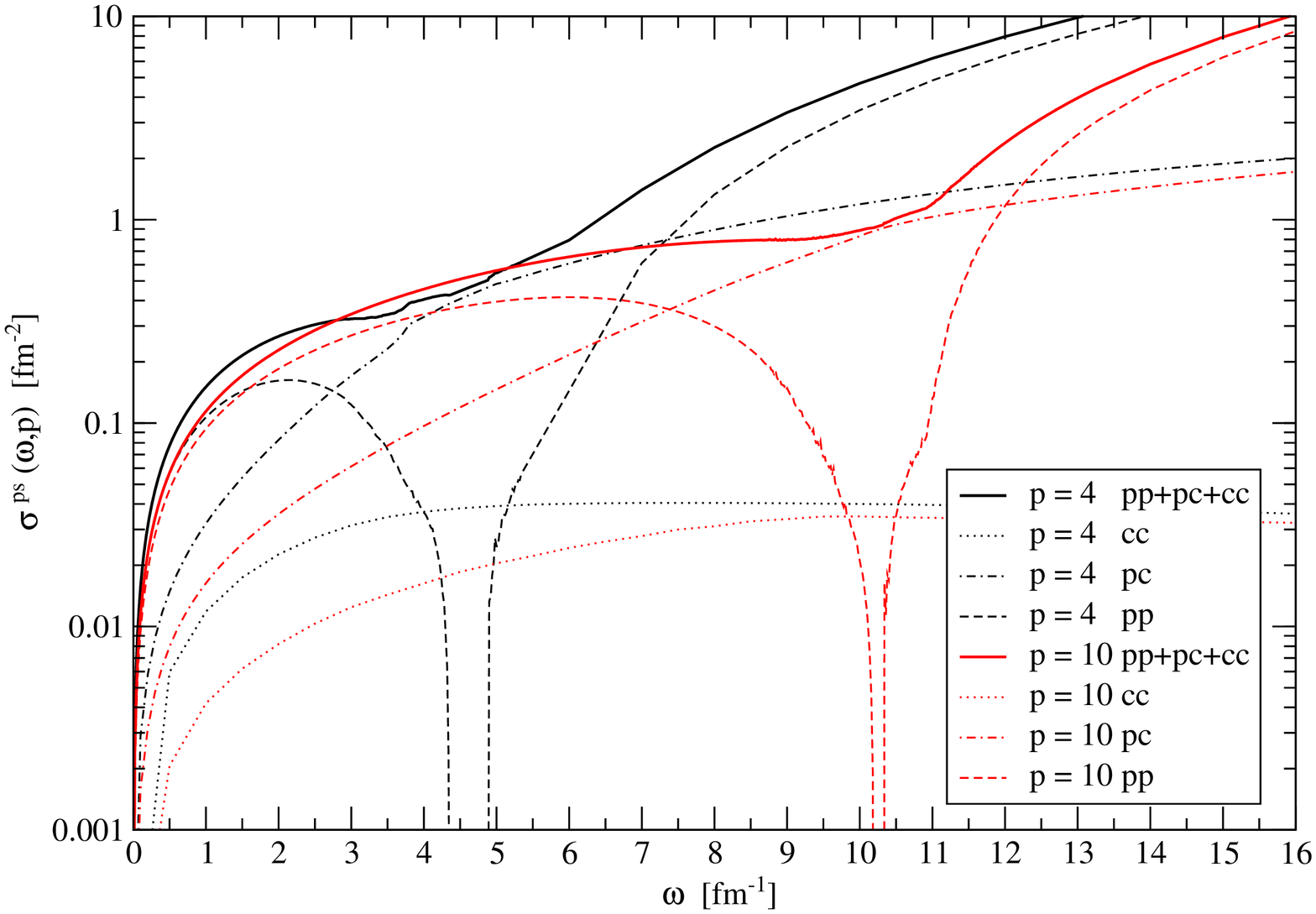}
\caption{The different contributions (pp,pc and cc) to the finite momentum pseudoscalar MSF for $p=4$  fm$^{-1}$ and $p=10$  fm$^{-1}$, at $T=2T_c$.}
\label{Fig:contib2tc} 
\end{center}
\end{figure}
\begin{figure}[!htp]
\begin{center}
\includegraphics[clip,width=0.70\textwidth,angle=270]{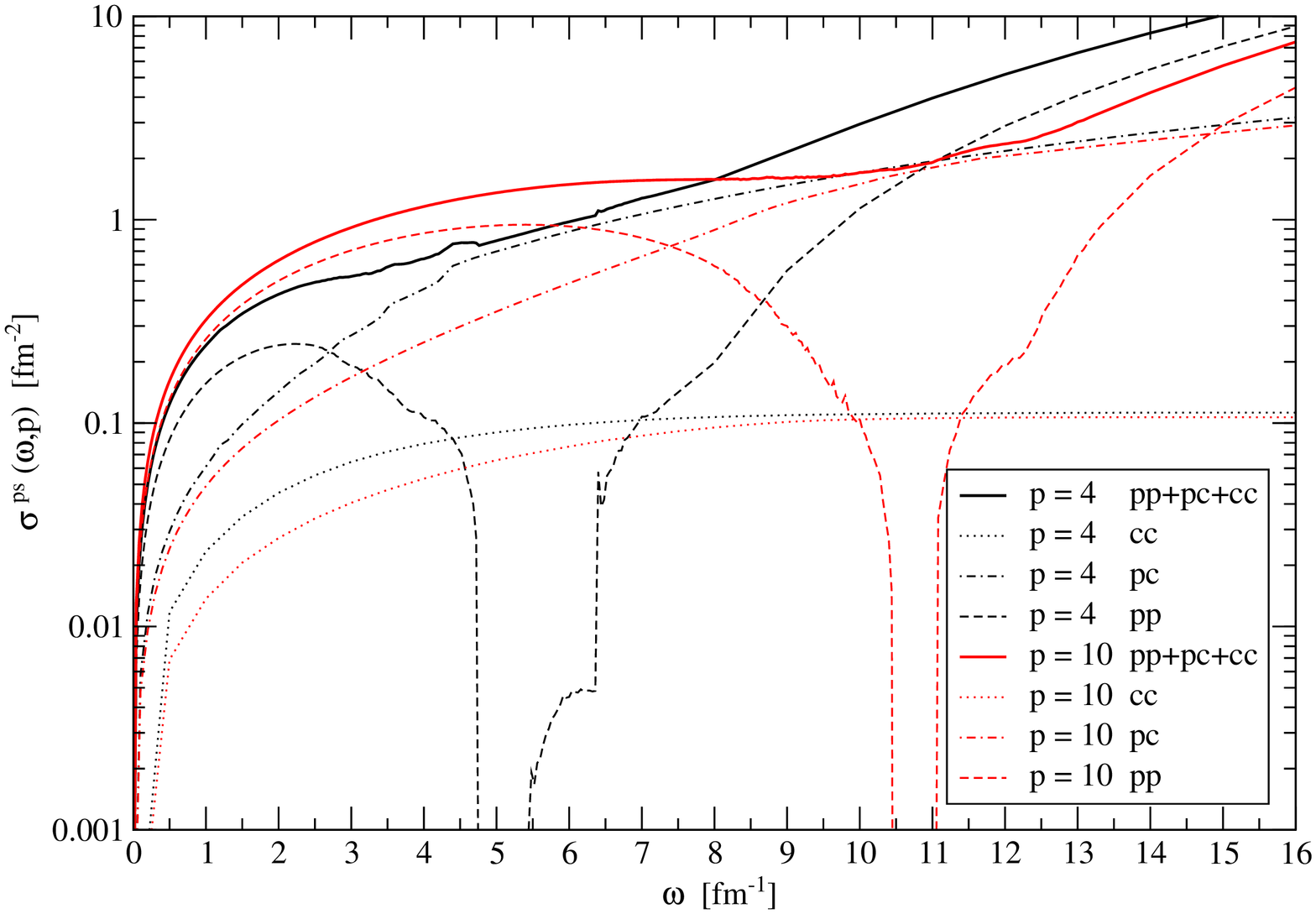}
\caption{The same as in Fig. \ref{Fig:contib2tc}, at $T=4T_c$.}
\label{Fig:contib4tc} 
\end{center}
\end{figure}
In Figs. \ref{Fig:contib2tc} and \ref{Fig:contib4tc} we show the different contributions [pole-pole (pp), pole-cut (pc) and cut-cut (cc)] to the finite momentum MSF for two different temperatures. It clearly appears that, for large enough frequencies, the pp term plays the dominant role: hence a closer inspection to the latter is mandatory.

\begin{figure}[!htp]
\begin{center}
\includegraphics[clip,width=0.70\textwidth,angle=270]{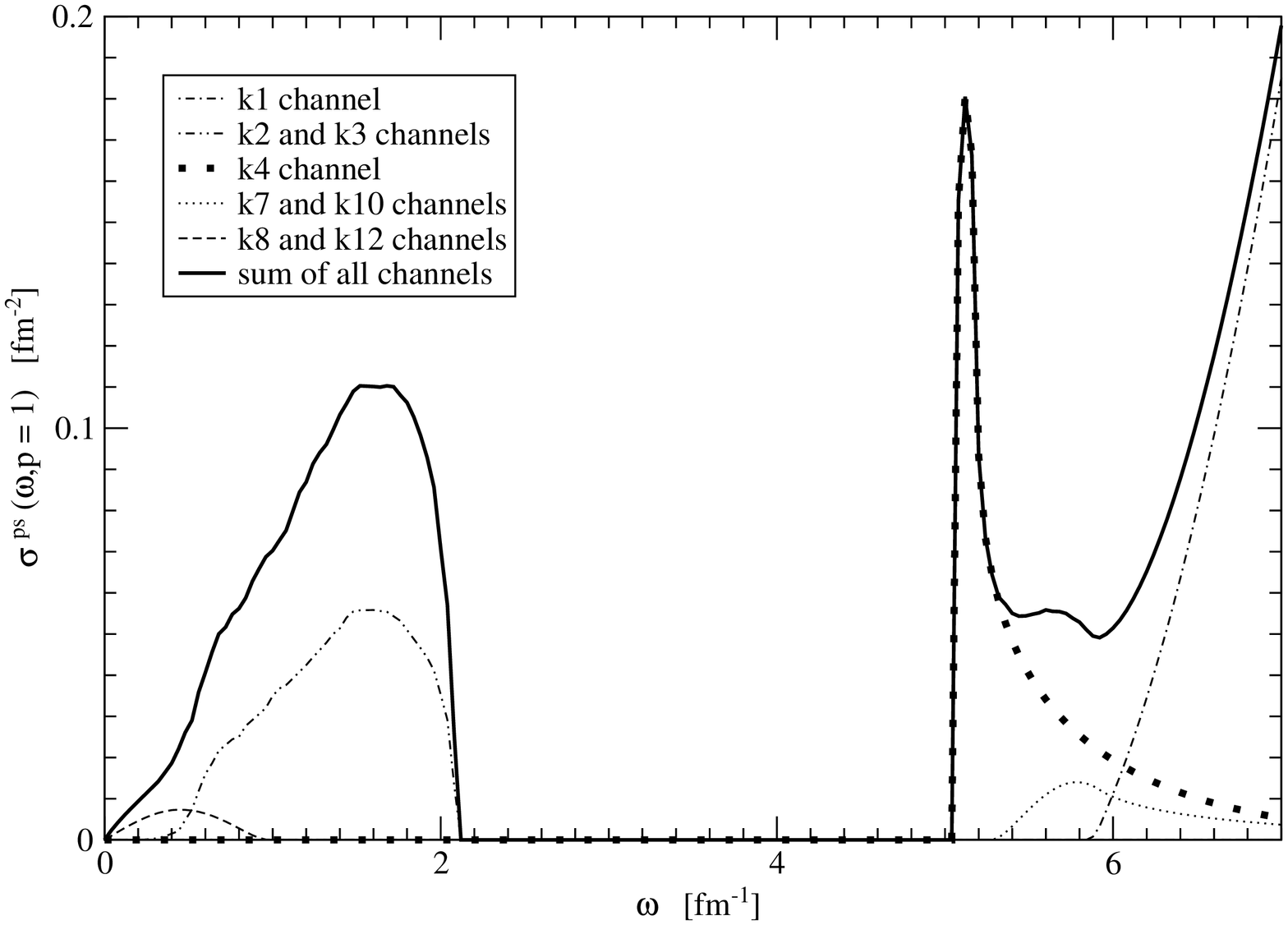}
\caption{The major processes contributing to the pole-pole term of the HTL MSF. The plot is given for the case $p=1$ fm$^-1$ and $T=2T_c$.}
\label{Fig:zoom1}
\end{center}
\end{figure}
\begin{figure}[!htp]
\begin{center}
\includegraphics[clip,width=0.38\textwidth,angle=270]{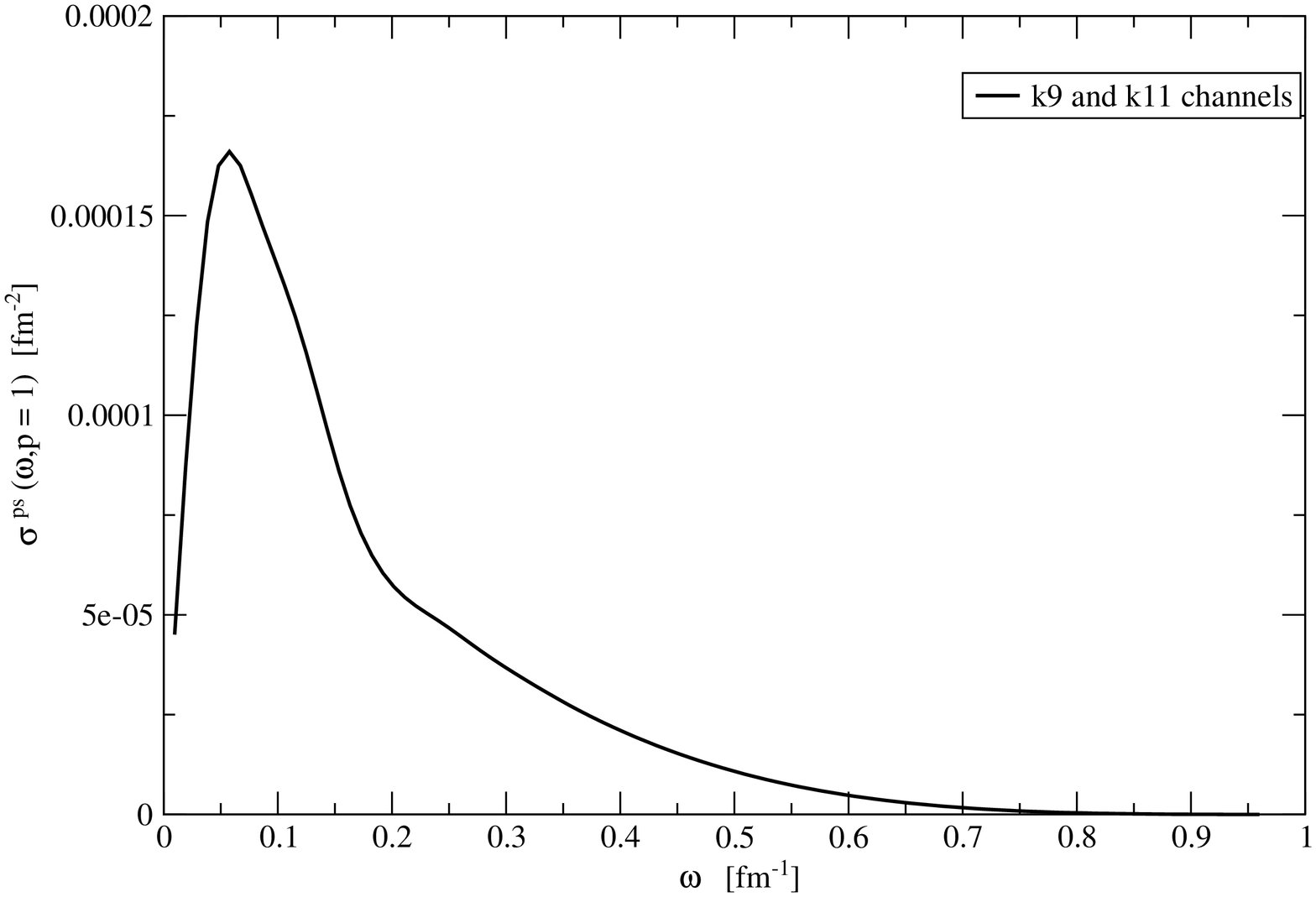}
\includegraphics[clip,width=0.38\textwidth,angle=270]{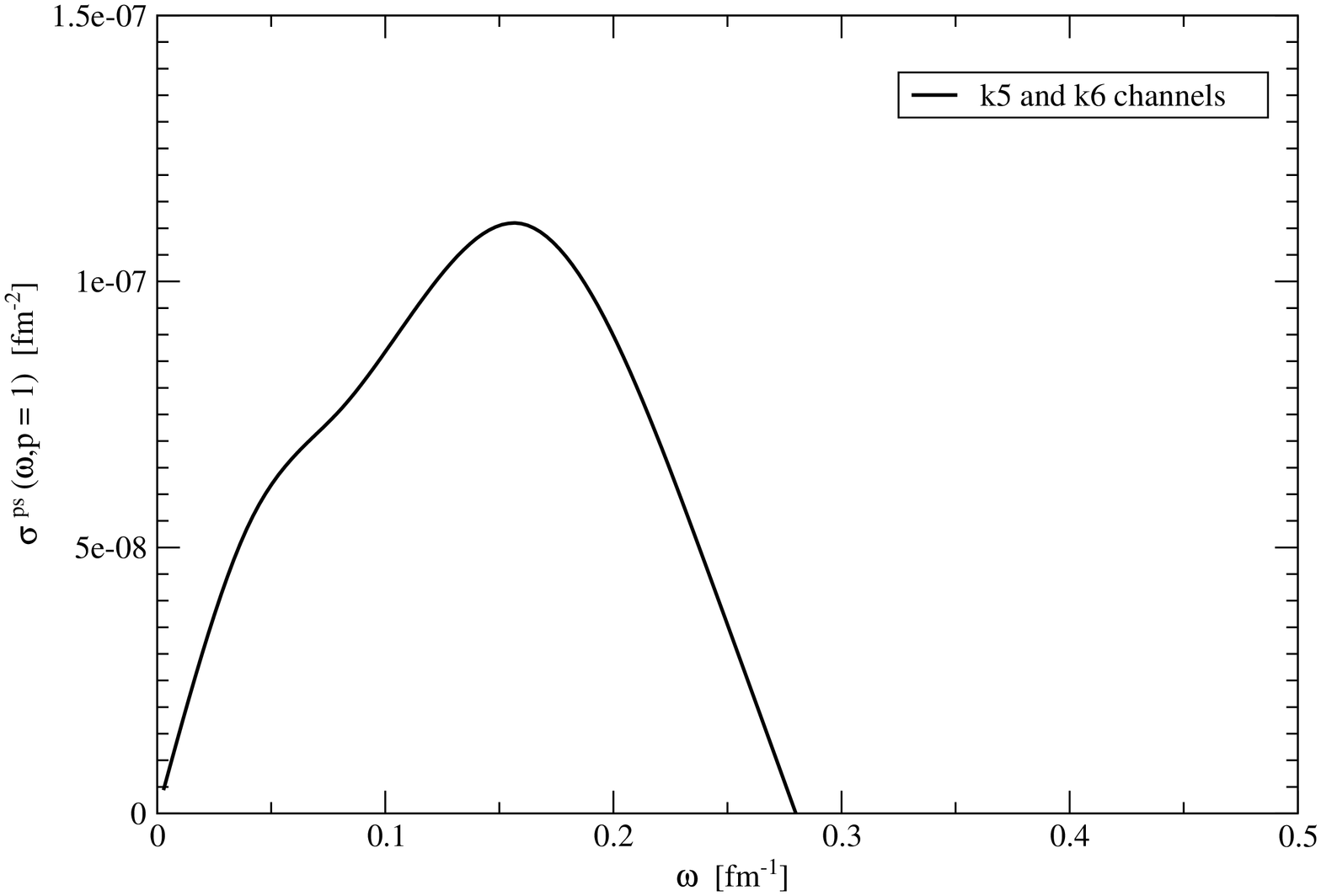}
\caption{The other contributions to the pole-pole HTL MSF for $p=1$ fm$^-1$ and $T=2T_c$.}
\label{Fig:zoom2} 
\end{center}
\end{figure}
Accordingly in Figs. \ref{Fig:zoom1} and \ref{Fig:zoom2} we display how the different processes quoted in Eq. (\ref{eq:processes}) contribute, at a given value of the temperature and of the spatial momentum, to the pole-pole term of the HTL MSF. In agreement with charge conjugation symmetry, it turns out that the processes 2-3, 7-10, 8-12, 9-11 and 5-6 give the same numerical contribution.\\
In analogy to the zero-momentum case studied in \cite{berry,mus}, it turns out that the dominant process at low energy (at zero momentum in fact it's the only one) is the decay $q_+\rightarrow q_-+M$ (together with its charge conjugate one); then the pp term displays a wide gap for an intermediate range of energy till when the plasmino-antiplasmino annihilation starts contributing. Such a process initially gives a quite large contribution due to the large density of states; then it decreases rapidly with the energy because of the very small value of the plasmino residue.\\
From Fig. \ref{Fig:zoom1} it appears that, for large enough frequencies, the dominant role is played by the quark-antiquark annihilation. Such a process starts contributing for $\omega$ larger than a threshold depending on the thermal gap mass $m_q$ acquired by the quarks in the thermal bath.\\
On the other hand the processes 9-11 and 5-6 ($\bar{q}_- \rightarrow  \bar{q}_- + M$ and $q_-\rightarrow q_+ + M$ together with their charge conjugate ones) turn out to be totally negligible, due to the low value of the plasmino residue and to the very small available phase space.
\subsection{Finite momentum temporal correlator}\label{sec:tempor}
In this Subsection we present our findings for the finite momentum temporal correlator defined in Eq. (\ref{eq:gtau}).\\
In order to assess the impact of the interaction it is convenient to consider the ratio
\beq
\frac{G_\trm{HTL}(-i\tau,\bold{p})}{G_\trm{free}(-i\tau,\bold{p})}=\frac{\displaystyle{
\int\limits_0^{+\infty}d\omega~\sigma_\trm{HTL}(\omega,\bold{p})K(\omega,\tau)}}{\displaystyle{
\int\limits_0^{+\infty}d\omega~\sigma_\trm{free}(\omega,\bold{p})K(\omega,\tau)}}\;.
\eeq
Such a ratio turns out to be finite for every value of $\tau$. In particular in the previous Subsection we showed how the non-interacting and the HTL MSF display the same high energy asymptotic behavior, both at zero and at finite spatial momentum. This implies that the above ratio approaches 1 for $\tau\to 0$ (or $\beta$). In fact, due to the structure of the thermal kernel given in Eq. (\ref{eq:gtau}), in this limit the correlator $G(\tau)$ is dominated by the high energy behavior of the MSF.   

\begin{figure}[!htp]
\begin{center}
\includegraphics[clip,width=0.60\textwidth,angle=270]{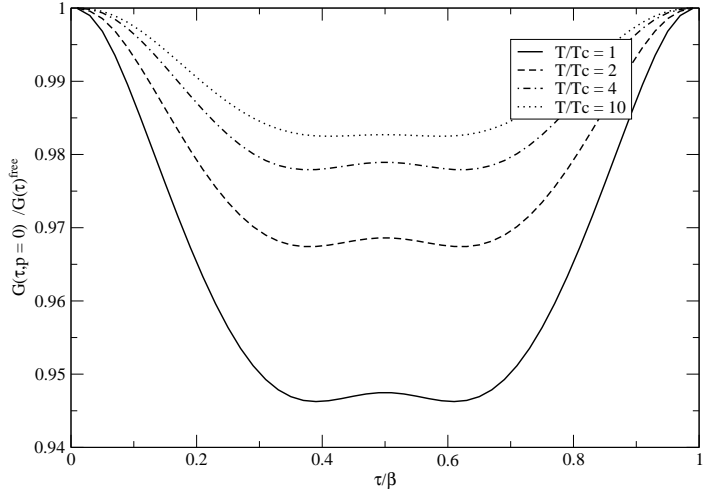}
\caption{The ratio $G_\trm{HTL}(\tau)/G_\trm{free}(\tau)$ for different temperatures at $p=0$ fm$^{-1}$.}
\label{Fig:htl_free0} 
\end{center}
\end{figure}
In Fig. \ref{Fig:htl_free0} we start by considering the zero momentum case (already studied in \cite{berry}), showing the ratio $G_\trm{HTL}(\tau)/G_\trm{free}(\tau)$ for a range of temperatures from $T=T_c$ to $T=10T_c$.

\begin{figure}[!htp]
\begin{center}
\includegraphics[clip,width=0.60\textwidth,angle=270]{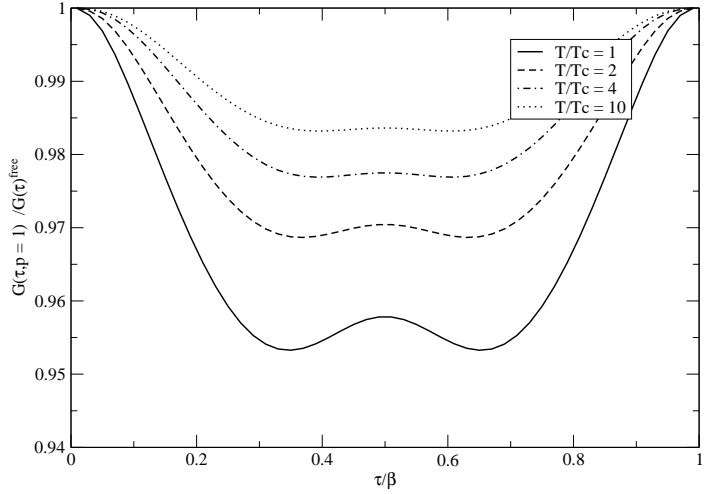}
\caption{The ratio $G_\trm{HTL}(\tau)/G_\trm{free}(\tau)$ for different temperatures at $p=1$ fm$^{-1}$.}
\label{Fig:htl_free1} 
\end{center}
\end{figure}
\begin{figure}[!htp]
\begin{center}
\includegraphics[clip,width=0.60\textwidth,angle=270]{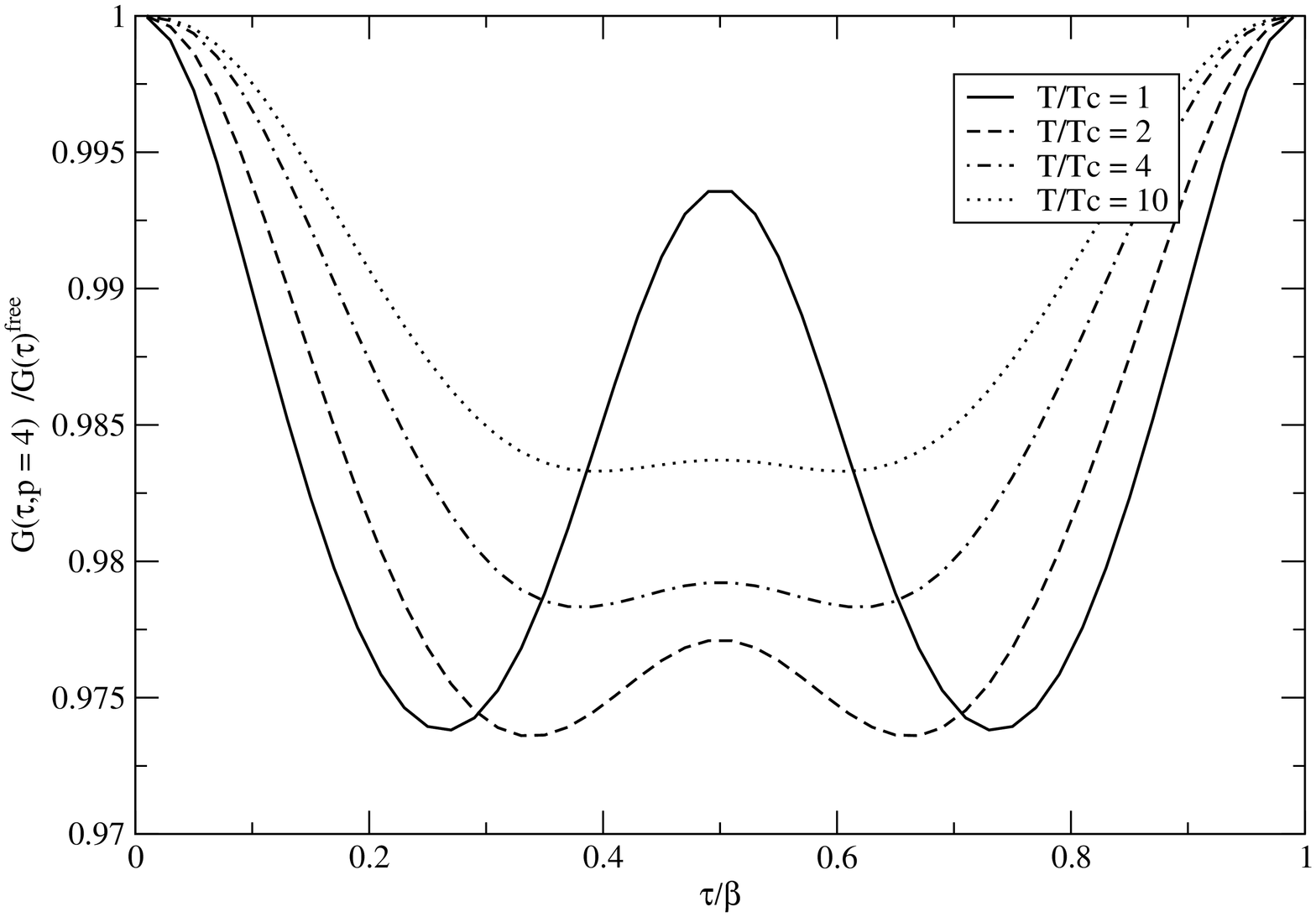}
\caption{The ratio $G_\trm{HTL}(\tau)/G_\trm{free}(\tau)$ for different temperatures at $p=4$ fm$^{-1}$.}
\label{Fig:htl_free4} 
\end{center}
\end{figure}
Then we move to the finite momentum case. In Figs. \ref{Fig:htl_free1} and \ref{Fig:htl_free4} the spatial momentum is kept fixed to the values $p=1$ fm$^{-1}$ and $p=4$ fm$^{-1}$, respectively, and the temperature is let to change in the range from $T=T_c$ to $T=10T_c$. As the temperature increases the ratio $G_\trm{HTL}(\tau)/G_\trm{free}(\tau)$ moves closer to $1$, reflecting the running of the coupling, and the bump at $\tau=\beta/2$ is smeared out.

\begin{figure}[!htp]
\begin{center}
\includegraphics[clip,width=0.50\textwidth,angle=270]{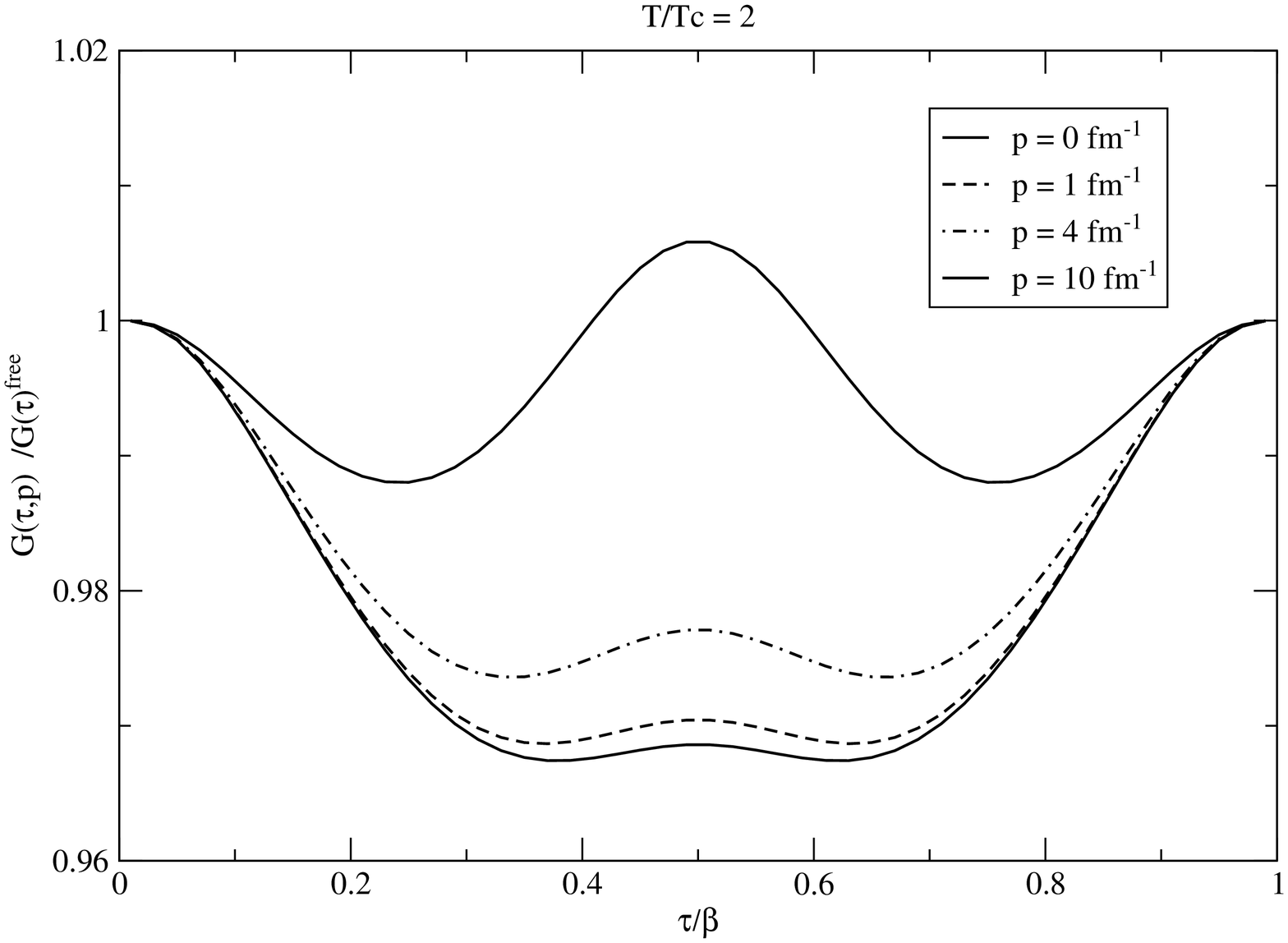}
\includegraphics[clip,width=0.50\textwidth,angle=270]{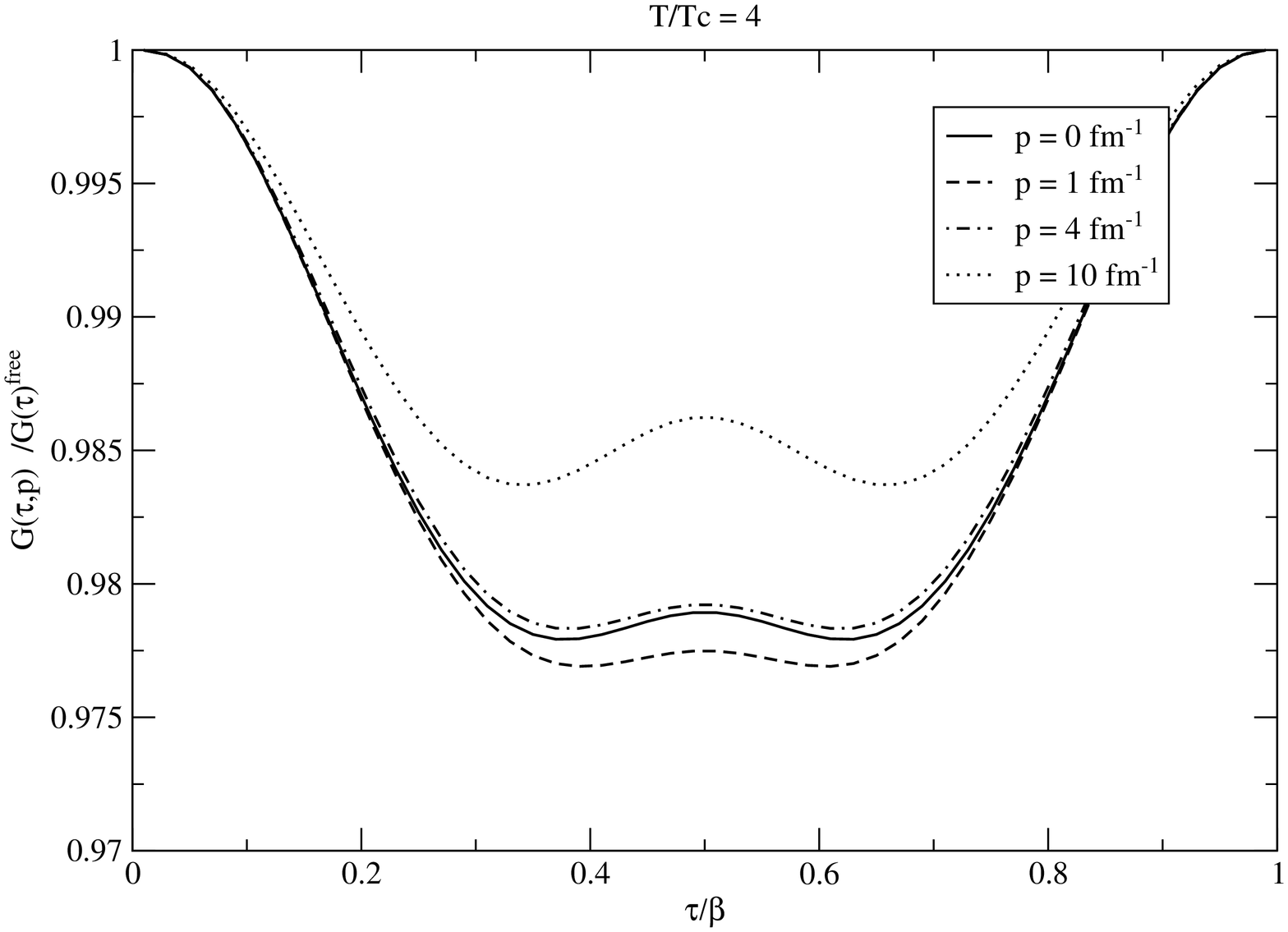}
\includegraphics[clip,width=0.50\textwidth,angle=270]{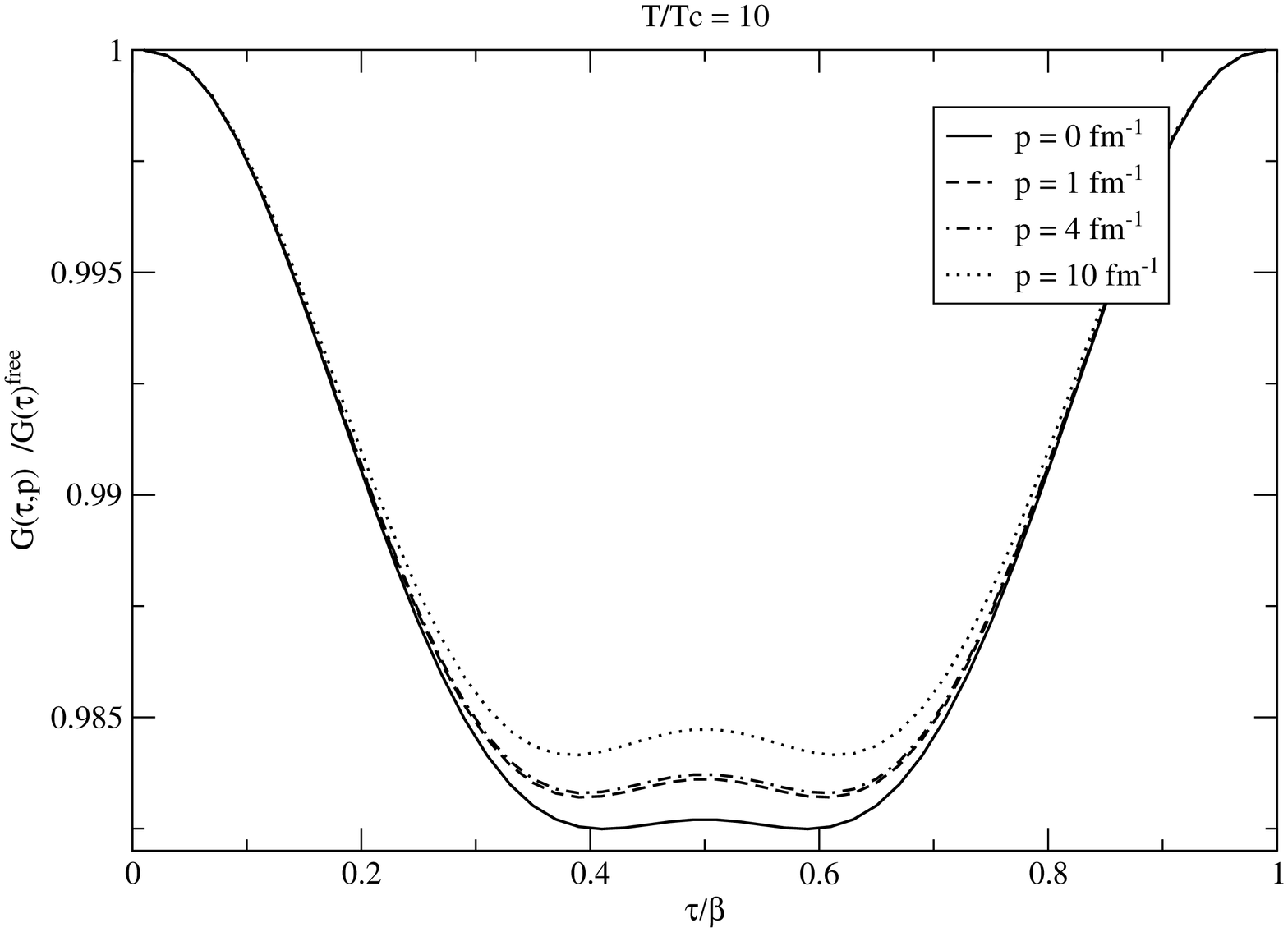}
\caption{The ratio $G_\trm{HTL}(\tau)/G_\trm{free}(\tau)$ for different values of the spatial momentum at $T=2T_c$, $T=4T_c$ and $T=10T_c$.}
\label{Fig:htl_freet2410} 
\end{center}
\end{figure}
In Fig. \ref{Fig:htl_freet2410} the temperature is kept fixed at the values $T=2T_c$, $T=4T_c$ and $T=10T_c$, respectively, and we display how the ratio $G_\trm{HTL}(\tau)/G_\trm{free}(\tau)$ is modified by changing the spatial momentum. The finite momentum effects are less and less important as the temperature increases, since the ratio $p/T$ gets smaller and smaller.

\begin{figure}[!htp]
\begin{center}
\includegraphics[clip,width=0.38\textwidth,angle=270]{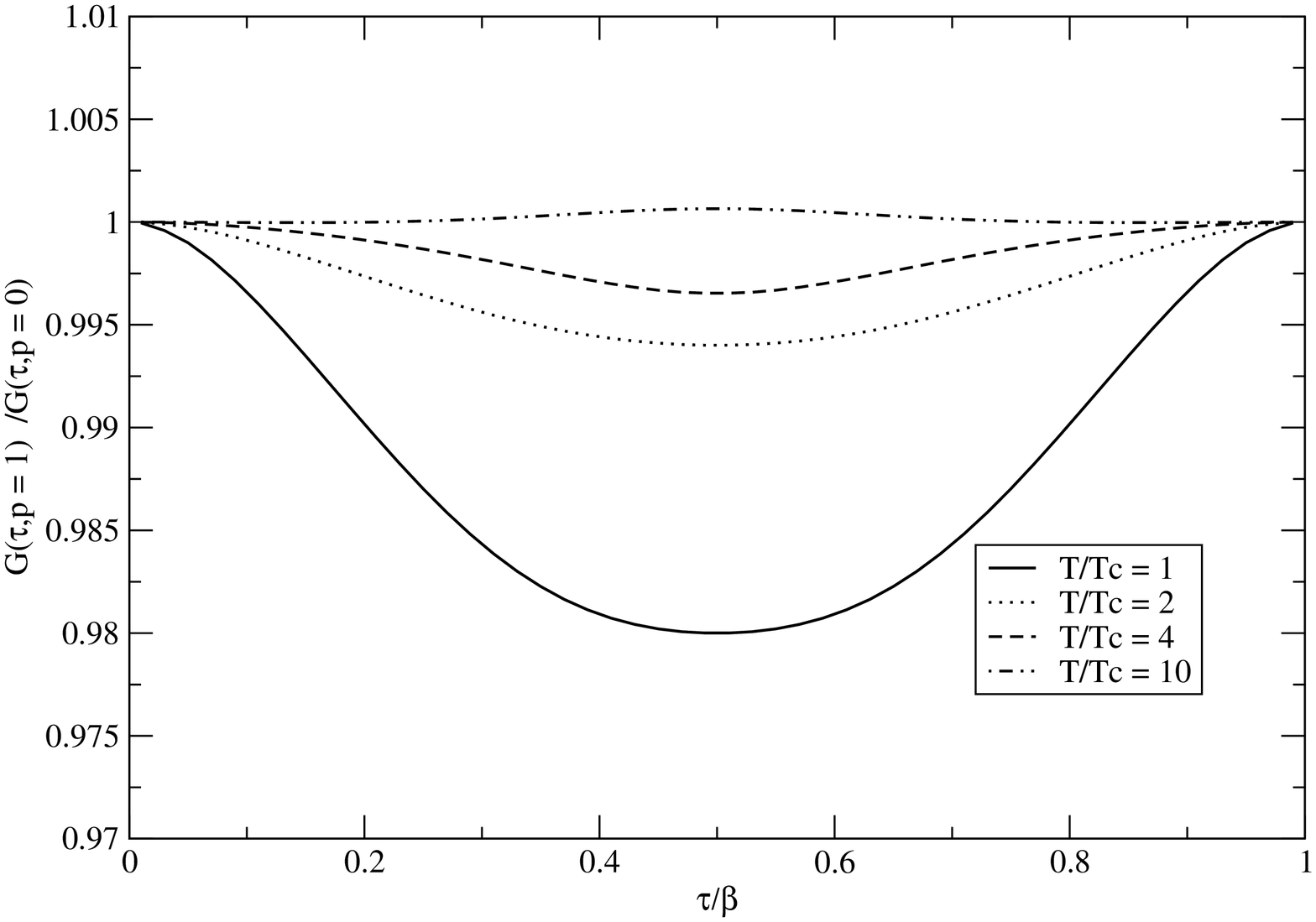}
\includegraphics[clip,width=0.38\textwidth,angle=270]{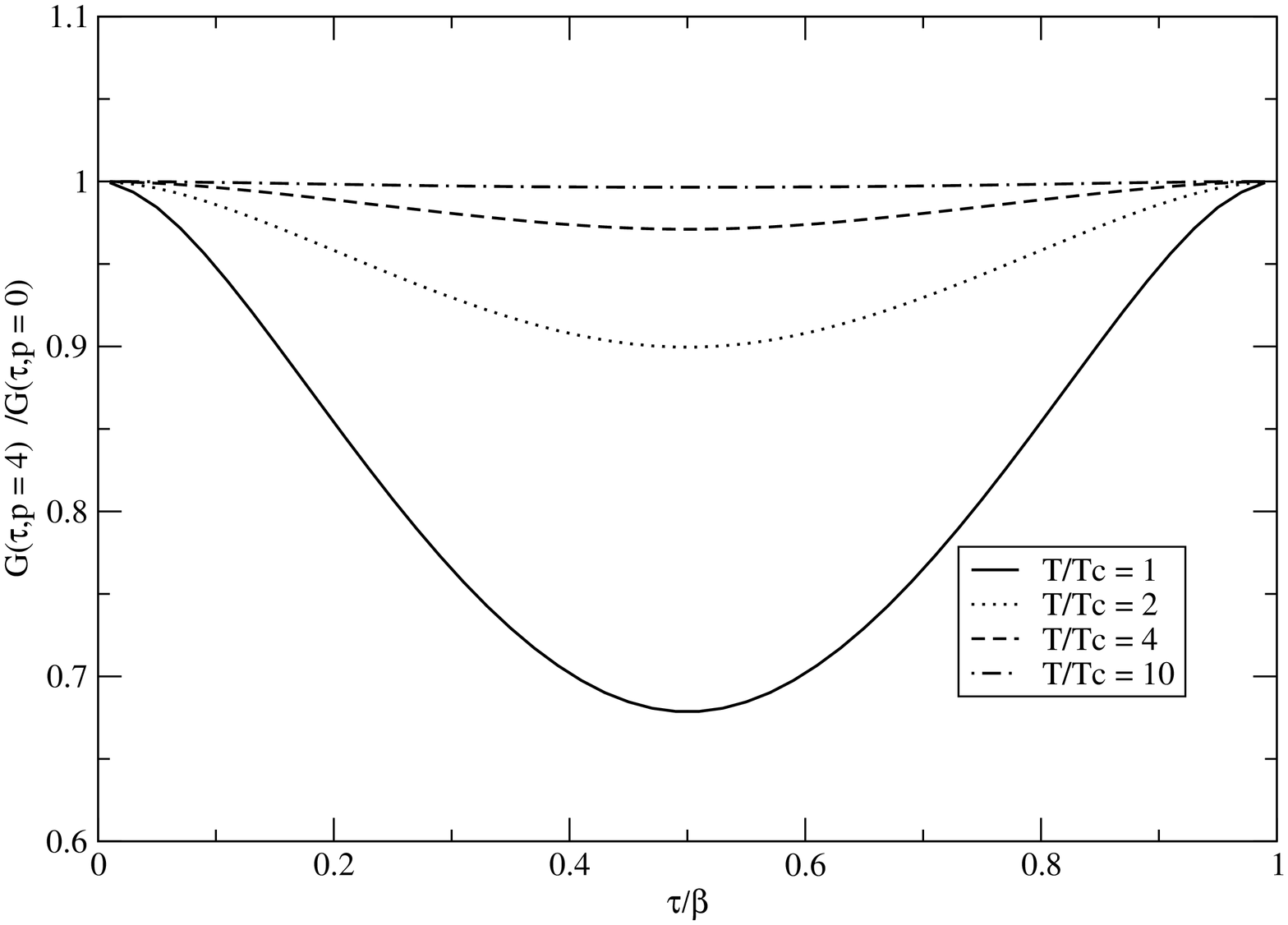}
\caption{The ratio $G_\trm{HTL}(\tau,p)/G_\trm{HTL}(\tau,p=0)$ for different temperatures at $p=1$ fm$^{-1}$ (left) and $p=4$ fm$^{-1}$ (right).}
\label{Fig:htlp14_htlp0} 
\end{center}
\end{figure}

\begin{figure}[!htp]
\begin{center}
\includegraphics[clip,width=0.38\textwidth,angle=270]{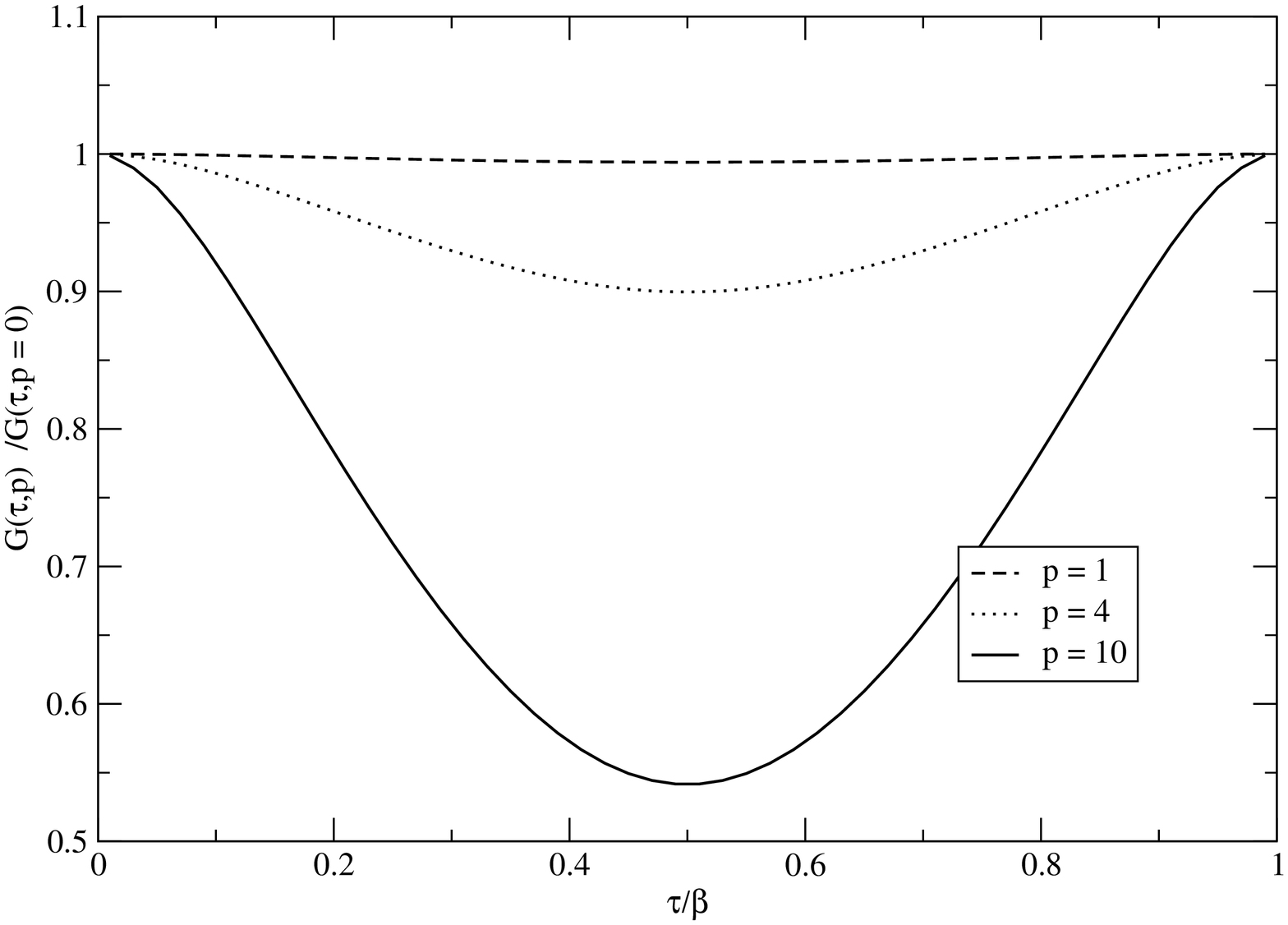}
\includegraphics[clip,width=0.38\textwidth,angle=270]{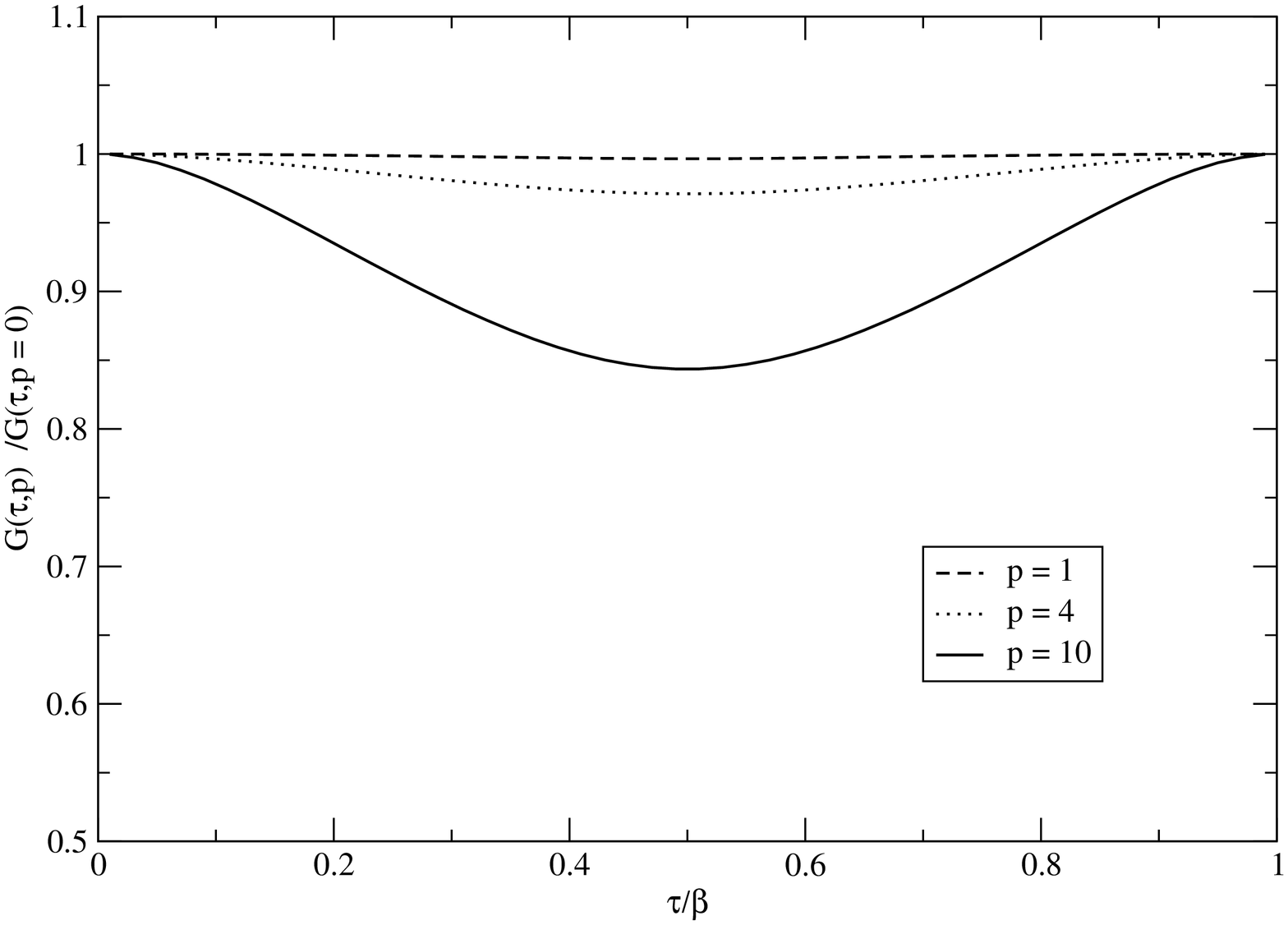}
\caption{The ratio $G_\trm{HTL}(\tau,p)/G_\trm{HTL}(\tau,p=0)$ for different values of the spatial momentum (measured in fm$^{-1}$) for $T=2T_c$ (left) and $T=4T_c$ (right).}
\label{Fig:htlp_htlp0_t24} 
\end{center}
\end{figure}

The role of the not vanishing spatial momentum in the interacting theory can be also made visible by considering the ratio
\beq
\frac{G_\trm{HTL}(-i\tau,\bold{p})}{G_\trm{HTL}(-i\tau,\bold{0})}=\frac{\displaystyle{
\int\limits_0^{+\infty}d\omega~\sigma_\trm{HTL}(\omega,\bold{p})K(\omega,\tau)}}{\displaystyle{
\int\limits_0^{+\infty}d\omega~\sigma_\trm{HTL}(\omega,\bold{0})K(\omega,\tau)}}\;,
\eeq 
which is displayed in Figs. \ref{Fig:htlp14_htlp0} and \ref{Fig:htlp_htlp0_t24}.
It is clearly apparent the substantial quenching of the HTL meson propagator (with respect to the zero momentum result) occurring, for large spatial momenta, around $\tau=\beta/2$.
\section{Conclusions}\label{sec:concl}
We have examined the impact of finite values of the spatial momentum on the spectral density and on the temporal correlation function of a pseudoscalar meson for temperatures above the deconfinement phase transition and zero chemical potential. This amounts to study the properties of an excitation carrying the quantum number of a meson and propagating (i.e. being not at rest) in the heat-bath frame.\\
In our treatment we employed HTL resummed quark propagators. This has allowed us to perform many (but, of course, not all!) calculations analytically and to identify the different physical processes contributing to the MSFs and to the temporal correlators, a task impossible to achieve, of course, when the same quantities are evaluated on the lattice (apart from the difficulties related to the MEM technique to extract a spectral density from an Euclidean correlator).\\
In our investigation we have explored a range of spatial momenta from 
$p=0$~fm$^{-1}$ to $p=10$~fm$^{-1}$ and of temperatures from $T=T_c$ (for the sake of completeness) to $T=10T_c$, although the HTL approximation is known to hold starting from temperatures of the order of $T\sim 2.5 T_c$. Clearly no intrinsic difficulty exists in extending our calculations below such a temperature, but the increase of the gauge coupling $g(T)$ as $T$ moves downward makes the separation of scales (hard and soft) which lies at the basis of the HTL approximation no longer warranted.\\

In summarizing our results we start by reminding the reader that, at zero momentum, the main features of the MSF are the presence of the Van Hove singularities (reflecting a divergence in the density of states) and the radically different low energy behavior of the HTL predictions with respect to the free case. Such a contrast results particularly visible in the plot of $\sigma(\omega,\bold{0})/\omega^2$, where, for $\omega\to 0$ the HTL curve diverges while the free result vanishes.\\
Moving to finite spatial momentum we first notice that many more terms, with respect to the zero momentum case, contribute to the MSF, reflecting the different physical processes which may occur and which have been discussed in detail in the text.\\
As a result of the calculation it turns out that the Van Hove singularities, so prominent in the zero momentum case, are smoothed out by the angular integration when $p\ne 0$.\\
Another finding worth to be pointed out is that while the free MSFs vanish on the light-cone, the HTL curves stay finite.\\
Finally the impact of a finite value of the spatial momentum has been investigated also for the HTL temporal correlator $G(\tau,\bold{p})$, for which we plotted the ratios with respect to the free and the zero momentum results.\\
In particular over the whole range of temperatures and momenta examined we found that the ratio of the HTL result with respect to the non-interacting correlator differs from one just for a few percent.\\
Hence, although we do not present here the Fourier transform (along the z-axis) of the finite momentum correlator (integrated over all the values of $\tau$), which indeed represent a major numerical effort (work is in progress in this direction) we do not expect dramatic differences with respect to the free result.\\  
We hope our work can provide a complementary and independent approach to the studies of MSFs and temporal correlators performed on the lattice.
\section*{Acknowledgments}
One of the authors (A.B.) gratefully acknowledges the financial support of the Della Riccia foundation. He also thanks the Institutions CEA-SPhT (Saclay) and ECT* (Trento) for the warm hospitality offered to him when this work was completed.\\
One of the authors (P.C.) thanks the Department of Theoretical Physics of the Torino University for the warm hospitality in the initial phase of this work.
\appendix
\section{Adjustment of the parameters}
For the $N_f=2$ transition temperature, following \cite{kacz}, we adopt the value $T_c=202$ MeV. The ratio $T_c/\Lambda_{\overline{MS}} = 0.7721$ for the $N_f=2$ case is taken from Refs. \cite{karsch,goc,zan}.\\
The running of the gauge coupling is given by the two-loop perturbative beta-function, leading to the expression:
\beq
g^{-2}(T) = 2 b_0 \log{\frac{\mu}{\Lambda_{\overline{MS}}}} +
\frac{b_1}{b_0} \log\Big\{2 \log{\frac{\mu}{\Lambda_{\overline{MS}}}}
\Big\}\label{eq:running}
\eeq
where
\beqa
b_0 &=& \frac{1}{16 \pi^2} \Big( 11 - 2 \frac{N_f}{3} \Big)
\nonumber \\
b_1 &=& \frac{1}{(16 \pi^2)^2} \Big( 102 - 38 \frac{N_f}{3} \Big)
\eeqa
The choice of the renormalization scale $\mu$ should reflect the typical momentum exchanged by the particles which, in an ultra-relativistic plasma, is of order $T$. For what concerns the precise numerical coefficient one should in principle let it vary within a reasonable range in order to get an estimate on the theoretical uncertainty of the calculation. Here, for the sake of simplicity, we simply adopt the choice $\mu = 1.142 \pi T$ suggested in \cite{kacz}.\\
We note that the running coupling $g(T)$ defined in Eq. (\ref{eq:running}) enters our results only through the thermal gap mass of the quark  $m_q=g(T) T/\sqrt{6}$.


\begin{thebibliography}{99}
\bibitem{berry} W.M. Alberico, A. Beraudo and A. Molinari,
                Nucl. Phys. {\bf A750} (2005), 359;
\bibitem{mus} F. Karsch, M.G. Mustafa, M.H. Thoma,
              Phys. Lett. {\bf B497} (2001), 249; 
\bibitem{bra} E. Braaten, R.D. Pisarski and T.C. Yuan,
              Phys. Rev. Lett. {\bf 64} (1990), 2242;
\bibitem{flor} W. Florkowski and B.L. Friman
               Z. Phys. {\bf A347} (1994), 271;
\bibitem{za} T.H. Hansson and I. Zahed,
             Nucl. Phys. {\bf B374} (1992), 277;
\bibitem{lai} M. Laine and  M. Vepsalainen,
              JHEP 0402 (2004) 004.
\bibitem{taro} QCD-TARO Collaboration (P. de Forcrand et al.),
               hep-lat/9901017;
\bibitem{taro2} QCD-TARO Collaboration (I. Pushkina et al.),
                Phys. Lett. {\bf B609} (2005), 265;
\bibitem{petr} P. Petreczky,
               J. Phys. {\bf G30} (2004), S431;
\bibitem{wis} S. Wissel et al.,
              hep-lat/0510031; 
\bibitem{aar} G. Aarts and J.M. Martinez Resco,
              Nucl. Phys. {\bf B726} (2005), 93;
\bibitem{aar2} G. Aarts, S. Hands, S.Y. Kim and J.M. Martinez Resco,
               PoS LAT2005 (2005), 182, hep-lat/0509062;
\bibitem{mem} M. Asakawa, T. Hatsuda and Nakahara,
              Prog. Part. Nucl. Phys. {\bf 46} (2001), 459;
\bibitem{dilep} F.Karsch et al.,
               Phys. Lett. B {\bf 530} (2002), 147;
\bibitem{gupta} S. Gupta,
                Phys. Lett. B {\bf 597} (2004), 57;
\bibitem{naka} A. Nakamura and S. Sakai,
               Phys. Rev. Lett. {\bf 94} (2005), 072305;
\bibitem{aar3} G. Aarts and J.M. Martinez Resco,
               JHEP {\bf 0204} (2002), 053;
\bibitem{wyld} F.Karsch and H.W. Wyld,
               Phys. Rev. D {\bf 35} (1987), 2518;
\bibitem{kubo1} R. Horsley and W. Schoenmaker,
                Nucl. Phys. B {\bf 280} (1987), 735; 
\bibitem{kubo2} D.N. Zubarev,
                Nonequilibrium Statistical Thermodynamic, Plenum, New York, 1974,
\bibitem{kubo3} A. Hosoya, M. Sakagami and M. Takao,
                Ann. Phys. {\bf 154} (1984), 229;
\bibitem{dat} S. Datta \textit{et al.},
              hep-lat/0409147;
\bibitem{lb} M. Le Bellac,
             Thermal Field Theory, Cambridge University Press, 1996;
\bibitem{bi} J.P. Blaizot and E. Iancu,
             Phys. Rept. {\bf 359} (2002), 355;
\bibitem{tho} M.H. Thoma,
              Z.Phys. C {\bf 66} (1995), 491;
\bibitem{bie} J.P. Blaizot and E. Iancu,
              Phys. Rev. D {\bf 63} (2001);
\bibitem{blasu} J.P. Blaizot, E. Iancu and A. Rebhan,
                Phys. Lett. B {\bf 523} (2001), 143;
\bibitem{blater} J.P. Blaizot, E. Iancu and A. Rebhan,
                 Phys. Lett. B {\bf 470} (1999), 181;
\bibitem{kacz} O. Kaczmarek and F. Zantow,
               Phys. Rev.{\bf D71} (2005), 114510;
\bibitem{karsch} F. Karsch, E. Laermann, and A. Peikert,
                 Phys. Lett. {\bf B478} (2000), 447.  
\bibitem{goc} M. Gockeler {\em et al.},
              Phys.Rev. {\bf D73} (2006) 014513,
\bibitem{zan} O. Kaczmarek and F. Zantow,
              hep-lat/0512031.
\end{thebibliography}
\end{document}